\documentclass[twocolumn,superscriptaddress,longbibliography,aps,pra,preprintnumbers]{revtex4-1}

\usepackage{amsmath,amssymb,amsfonts}
\usepackage{graphicx}
\usepackage{dcolumn}
\usepackage{bm}
\usepackage{epstopdf}
\usepackage{natbib}

\usepackage[urlcolor=blue,colorlinks=true,citecolor=blue,linkcolor=blue,pdfstartview={FitH},bookmarks=false]{hyperref}

\usepackage[T1]{fontenc}
\usepackage{lmodern}

\begin{document}

\preprint{Submitted to: Physical Review Materials}

\title{Superconducting monolayer deposited on substrate: effects of  the spin-orbit coupling induced by proximity effects}

\author{Andrzej Ptok}
\email[e-mail: ]{aptok@mmj.pl}
\affiliation{Institute of Nuclear Physics, Polish Academy of Sciences, ulica W. E. Radzikowskiego 152, PL-31342 Krak\'ow, Poland}
\affiliation{Institute of Physics, Maria Curie-Sk\l{}odowska University, Plac M. Sk\l{}odowskiej-Curie 1, PL-20031 Lublin, Poland}

\author{Karen Rodr\'{i}guez}
\email[e-mail: ]{karem.c.rodriguez@correounivalle.edu.co}
\affiliation{Departamento de F\'isica, Universidad del Valle, A. A. 25360, Cali, Colombia}
\affiliation{Centre for Bioinformatics and Photonics -- CiBioFi, Calle 13 No. 100-00, Edificio 320 No. 1069, Cali, Colombia}

\author{Konrad Jerzy Kapcia}
\email[e-mail: ]{konrad.kapcia@ifj.edu.pl}
\affiliation{Institute of Nuclear Physics, Polish Academy of Sciences, ulica W. E. Radzikowskiego 152, PL-31342 Krak\'ow, Poland}
\affiliation{Institute of Physics, Polish Academy of Sciences, Aleja Lotnik\'{o}w 32/46, PL-02668 Warsaw, Poland}

\date{\today}

\begin{abstract}
The spin-orbit coupling can lead to exotic states of matter and unexpected behavior of the system properties.
In this paper, we investigate the influence of spin-orbit coupling induced by proximity effects on a monolayer of superconductor (with {\it s-wave} or {\it d-wave} pairing) placed on an insulating bulk.
We show that the critical temperatures $T_{c}$ of the superconducting states can be tuned by the spin-orbit coupling both in the case of on-site and inter-site pairing.
Moreover, we discuss a possibility of changing the location of the maximal $T_{c}$ from the half-filling into the underdoped or overdoped regimes.
\end{abstract}


\maketitle

\section{Introduction}

The proximity effects may occur in a situation, when two different metals are brought together~\cite{degennes.64}.
In fact, the proximity effects can be treated as a {\it leakage} of some physical properties or quantities from one to another material.
As a consequence, for a material that does not posses particular features, a contact with a source material is sufficient enough to acquire them.
For the first time it was observed in a superconductor/normal metal/superconductor (S/N/S) junction~\cite{holm.meissner.32}.
The proximity effect is also experimentally observed as a supercurrent inside an insulator placed between two superconducting materials~\cite{anderson.rowell.63} -- a well-known Josephson effect~\cite{josephson.62}.

In modern physics, the proximity effects play an important role in many aspects of spintronics~\cite{zutic.fabian.04,bandyopadhyay.cahay.04,naber.feaz.07,bandyopadhyay.cahay.15},
where electron spins are exploited as an additional degree of freedom.
In this context, the possibility of manipulation of single spins by, e.g., inversion symmetry breaking effects is an important issue.
The spin-orbit (SO) coupling~\cite{dresselhaus.55,rashba.60,bychkow.rashba.84}, which mixes two spin directions~\cite{bansil.lin.16}, is an example of such an effect.
As a consequence, the spin is not longer a good quantum number.
These facts lead to several interesting phenomena, which can be applicable in spintronics devices~\cite{jungwirth.wunderlich.12,manchon.koo.15,joshi.16,soumyanarayanan.reyren.16}, e.g., as data storages~\cite{chappert.fert.07} or quantum computers~\cite{knill.05,joseph.eric.15,wang.kumar.16}.

Recently, a combination of the proximity effects and the SO coupling plays an important role in studying different types of junctions or heterostructures.
An example of that is the interplay between superconductivity and ferromagnetism (F), which can be experimentally investigated in S/F/S Josephson junctions.
That interplay leads to in- and out-of-plane magneto-anisotropies of the Josephson currents~\cite{linder.robinson.15,gingrich.niedzielski.16}, whose direction is controlled by the strength of the SO coupling~\cite{costa.hogl.17}.
In these heterostructures the interfacial Rashba SO coupling has been proposed as the mechanism from which the spin-flip Andreev reflection stems~\cite{niu.12}.

Similar behavior can be also helpful in realisation of the Majorana {\it quasi}-particles~\cite{alicea.12,beenakker.13,eliot.franz.15} in nanoobjects, such as quantum wires~\cite{kitaev.01}.
Recent experiments describe observation of the Majorana bound states in strongly SO coupled wires, which acquire superconductivity from proximity effects~\cite{mourik.zuo.12,das.ronen.12,
deng.yu.12,churchill.fatemi.13,feldman.randeria.17,deng.vaitiekenas.16,ptok.kobialka.17}.
Similar effects are expected in superconducting layers deposited on topological insulators.
In such systems, non-trivial topological states can be induced inside superconducting vortices~\cite{fu.kane.08,tewari.dassarma.07}.
The SO coupling induced in a superconductor through the proximity of strongly SO coupled topological insulator is also observed.

\paragraph*{Motivation.}
--- Examples presented above show a crucial role of proximity effects on the properties of the system.
Recent experiments revealed an extraordinary increase of the critical temperature of the the FeSe monolayer grown on (001) surface of SrTiO$_{3}$   from $8$ K~\cite{hsu.luo.08} exceeding to $65$ K~\cite{qingyan.zhi.12,liu.zhang.12,he.he.13,tan.zhang.13}.
Growth on the (110) surface of  SrTiO$_{3}$ is also possible and it gives an increase of the critical temperature up to $31.6$ K~\cite{zhou.zhang.16,zhang.peng.16,wu.dai.16}.
It should be mentioned that the surface of the  SrTiO$_{3}$ exhibits an effective spin-orbit effect~\cite{zhong.toth.13}.
It is generally agreed that mutual exchange of properties between surface and substrate leads to this unusual phenomena.
In this context, it  seems interesting to study a  possible effects of the {\it induced} SO coupling in superconducting monolayer, which does not show it initially.
In this case, changes of global physical properties of such a layer can lead to some novel and unexpected effects, as we show further in this work.

The qualitative process of the induction of an effective SO coupling by proximity effect can be explained in the following way.
The intrinsic SO coupling existing in the bulk substrate modifies wave functions of electrons located in the bulk.
A finite hybridization, arising from the overlapping orbitals of atoms in the substrate and the layer, leads to a modification of the band structure of the electrons belonging to the layer~\cite{wilson.nguyen.17}.
Effects of this modification of band structure can be described effectively as a spin-orbit coupling in the layer induced by proximity effects.

The presented idea of the induced SO coupling by proximity effects is realizable experimentally,
e.g. in the form of
Bi$_{2}$Te$_{3}$/Fe$_{1+y}$Te~\cite{he.shen.16},
WS$_{2}$/graphene~\cite{avsar.tan.14},
or Au/graphene~\cite{marchenko.varykhalov.12} heterostructures,
graphene at antiferromagnetic substrate~\cite{qiao.ren.14},
and carbon nanotubes coupled to a superconducting substrate~\cite{chudzinski.15}.
In these systems the proximity effects are crucial for the occurrence of the SO coupling.
Also some other modifications of the heterostructures can lead to the SO coupling, e.g. a change of the impurity structure.
These types of manipulations of the effective SO coupling based on the proximity effects is realised in non-magnetic/ferromagnetic bilayer~\cite{zhang.sun.15}.
However, also other possibilities of the {\it induced} SO coupling are studied.
A good example is the generation of the SO coupling in hydrogenated graphene~\cite{balakrishnan.kokwaikoon.13} or by the presence of impurities in graphene~\cite{castroneto.guinea.09}.

It should be mentioned that the situations described above are different from that where the SO exists in the whole volume of bulk material.
In such groups of systems  one can distinguish e.g. topological insulators~\cite{fu.kane.07,qi.zhang.11,bansil.lin.16} or topological superconductors~\cite{mackenzie.maeno.03,bergeret.volkov.05,
sato.fujimoto.09,smidman.salamon.17,sato.takahasi.10}.
In the latter case superconductivity exists in the presence of the SO, what effectively leads to the emergence of the {\it p-wave} gap symmetry from a conventional {\it s-wave} one~\cite{sato.takahasi.10,gorkov.rashba.01,zhang.tewari.08,alicae.10,
seo.han.12,yu.wu.16}.
Moreover, a coexistence of both phenomena is useful for a manipulation of the properties in different types of junctions~\cite{tanaka.yokoyama.09}.

In this work we analyze the effects of spin-orbit coupling on the critical temperature of a superconducting layer (with both {\it s-wave} and {\it d-wave} effective pairing) placed on the surface of an insulator.
We solve the effective model for the layer and determine critical temperatures as functions of the SO coupling and chemical potential.
It is shown that for fixed spin-orbit interaction the maximal critical temperature occurs in the system with optimal electron doping away from half-filling.

Next parts of this work are organized as follows.
In Section~\ref{sec.model}, we describe the model and method used, whereas in Section~\ref{sec.numerical} the numerical results are presented.
Section~\ref{sec.discuss} is devoted to the discussion of derived numerical results.
Finally, a summary and final comments are included in Section~\ref{sec.summary}.

\begin{figure}[!t]
\begin{center}
\includegraphics[width=\linewidth,keepaspectratio]{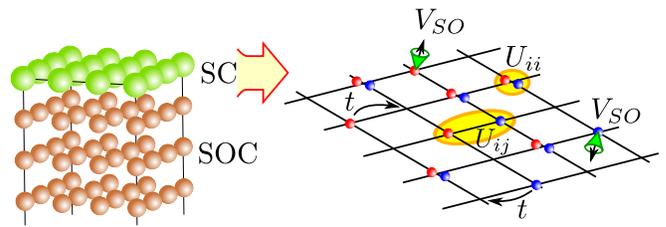}
\end{center}
\caption{Schematic representation of the analyzed interface (left) and its effective model (right).
We assume an existence of a two-dimensional superconducting (SC) monolayer placed on a surface of an insulator with a strong spin-orbit coupling (SOC).
We study the superconducting layer with half-spin particles (red and blue balls denotes opposite spins), which can move between sites of the lattice (e.g., between the nearest-neighbors with hopping amplitude $t$).
The on-site ($U_{ii}$) or inter-site ($U_{ij}$) attractive (negative) interactions (schematically indicated by yellow areas) are sources of a singlet pairing.
An effective spin-orbit coupling interaction ($V_{SO}$) in the layer is introduced by proximity effects from the bulk SOC material.
\label{fig.schem}
}
\end{figure}

\section{Model and method}
\label{sec.model}

For theoretical studies of the systems described in previous section we analyze the following model.
The considered system is schematically shown in Fig.~\ref{fig.schem} (the left panel).
The superconducting monolayer creates an interface with substrate with strong spin-orbit coupling (these two materials are denoted by SC and SOC, respectively).
Because of the proximity effects between SC and SOC, the SO coupling originating from the SOC material enters the SC layer.
We describe this scenario, emphasising the role of the induced SO coupling, by the effective model illustrated schematically in the right panel of Fig.~\ref{fig.schem}) and described in detail below (Eq.~(\ref{eq.hamrealspace})).
In our analyses we do not assume a type of superconductivity occurring in the material from which the monolayer is made.
Thus, we consider both on-site $U_{ii}$ and inter-site $U_{ij}$ pairing interactions (shown by yellow areas) between two electrons with opposite spin (singlet pairing).
This allows us to consider both {\it s-wave} and {\it d-wave} superconductors.
In the considered model, electrons with both directions of spin (represented by blue and red balls) can move in the superconducting plane between nearest-neighbor (NN) sites with hopping integral $t$ and next-nearest-neighbors (NNN) with hopping amplitude $t'$ (not shown in the schematic picture).

We describe our system by a tight-binding model with the Rashba-type SO interaction~\cite{li.covaci.11,li.covaci.12}.
The Hamiltonian acquires the form $\hat{H} = \hat{H}_{0} + \hat{H}_{I} + \hat{H}_{SO}$, where $\hat{H}_{0}$ denotes the non-interacting term (free electrons), $\hat{H}_{I}$ denotes the interaction between electron with opposite spins (source of superconductivity in the system), whereas $\hat{H}_{SO}$ describes the spin-orbit coupling.
The terms of the Hamiltonian have the following forms:
\begin{eqnarray}
\label{eq.hamrealspace}
\hat{H}_{0} &=& \sum_{ ij \alpha} \left( - t_{ij} - \mu \delta_{ij} \right) \hat{c}_{i\alpha}^{\dagger} \hat{c}_{j\alpha}, \nonumber \\
\hat{H}_{I} &=& \sum_{ij} U_{ij} \hat{c}_{i\uparrow}^{\dagger} \hat{c}_{i\downarrow} \hat{c}_{j\downarrow}^{\dagger} \hat{c}_{j\downarrow}, \\
\hat{H}_{SO} &=&
i V_{SO} \sum_{ i \alpha\beta } \left( \hat{c}_{i\alpha}^{\dagger} \sigma_{x}^{\alpha\beta} \hat{c}_{i+\hat{y} \beta} - \hat{c}_{i\alpha}^{\dagger} \sigma_{y}^{\alpha\beta} \hat{c}_{i+\hat{x}\beta} + \mbox{h.c.} \right) , \nonumber
\end{eqnarray}
where
$\hat{c}_{i\alpha}$ ($\hat{c}_{i\alpha}^{\dagger}$) is the annihilation (creation) operator of an electron at $i$-th site with spin $\alpha \in \{ \uparrow, \downarrow \}$,
$t_{ij}$ is the hopping integral between  $i$-th and $j$-th sites, $\mu$ is the chemical potential, and $U_{ij} < 0$ is the pairing interaction.
We consider both, the on-site interaction $U_{ij} = U \delta_{ij}$ corresponding to {\it s-wave}, and inter-site interaction between nearest neighbors $U_{ij} = U \left( \delta_{i\pm\hat{x},j} + \delta_{i\pm\hat{y},j} \right)$ corresponding to {\it d-wave} symmetry of the energy gap~\cite{ptok.crivelli.13}.
Here, $V_{SO}$ denotes the strength of the effective Rashba SO interaction induced by the proximity effects, while $\sigma^{\alpha \beta}_\tau$ is  $\alpha\beta$-component of the Pauli matrix $\check{\sigma}_\tau$ ($\tau \in \{ x,y \}$).
Finally, $\mu$ is the chemical potential, which determines the filling of the system.

In the momentum space, after employing the broken-symmetry Hartree-Fock mean-field approximation, the Hamiltonian terms (\ref{eq.hamrealspace}) are rewritten in the following forms:
\begin{eqnarray}
\label{eq.hammomspace}
\hat{H}_{0} &=& \sum_{ {\bm k} \alpha} \bar{E}_{{\bm k}\alpha} \hat{c}_{{\bm k}\alpha}^{\dagger} \hat{c}_{{\bm k}\alpha}, \nonumber  \\
\hat{H}_{I} &=& U \sum_{\bm k} \left( \Delta_{0} \gamma ( {\bm k} ) \hat{c}_{{\bm k}\uparrow}^{\dagger} \hat{c}_{-{\bm k}\downarrow}^{\dagger} + \Delta_{0}^{\ast} \gamma ( {\bm k} ) \hat{c}_{-{\bm k}\downarrow} \hat{c}_{{\bm k}\uparrow} \right), \nonumber\\
&-& U \sum_{\bm k} | \Delta_{0} |^{2} \gamma^{2} ( {\bm k} ) \\
\hat{H}_{SO} &=& \sum_{ {\bm k} \alpha \beta} \left(\check{V}_{\bm k}\right)_{\alpha\beta} \hat{c}_{{\bm k}\alpha}^{\dagger} \hat{c}_{{\bm k}\beta}, \nonumber
\end{eqnarray}
where $\bar{E}_{{\bm k}\alpha} = E_{\bm k} - \mu$ and $\left( \check{V}_{\bm k}\right)_{\alpha\beta}$ is $\alpha\beta$-component of $\check{V}_{\bm k} = 2 V_{SO} ( \sin ( k_{y}) \check{\sigma}_{x} - \sin ( k_{x} ) \check{\sigma}_{y} )$.
In the case of a square lattice with hopping between nearest neighbors, $t_{ij} = t \left( \delta_{i\pm\hat{x},j} + \delta_{i\pm\hat{y},j} \right)$, and next-nearest neighbors, $t_{ij} = t' \left( \delta_{i\pm(\hat{x}+\hat{y}),j} + \delta_{i\pm(\hat{x}-\hat{y}),j} \right)$, the dispersion relation is given by $E_{\bm k} = -2 t ( \cos (k_{x}) + \cos (k_{y}) ) - 4 t' \cos(k_{x}) \cos(k_{y})$.
The coefficient $\gamma( {\bm k} )$ describing the symmetry of the order parameter is either 1 or $\cos( k_{x} ) - \cos( k_{y} )$ for the {\it s-wave} and {\it d-wave} symmetry, respectively~\cite{halboth.walter.00,ptok.crivelli.15,ptok.crivelli.13}.
Finally, $\Delta_{0}=1/N \sum_{\bm k} \langle \hat{c}_{-{\bm k}\downarrow} \hat{c}_{{\bm k}\uparrow} \rangle$ is the amplitude of the superconducting order parameter, which is determined variationally by minimizing the grand canonical potential, cf. also Refs.~\cite{ptok.crivelli.13,ptok.14,micnas.ranninger.90}.
Notice that in Hamiltonian (\ref{eq.hammomspace}) we only left terms associated with (extended) BCS-type pairing -- the total momentum ${\bm Q}$ of the Cooper pair is zero: $| {\bm Q} |=0$~\cite{micnas.ranninger.90,ptok.crivelli.15,bardeen.cooper.57a,bardeen.cooper.57b}.
This assumption is valid only if the SO coupling in the monolayer is induced by proximity effects.

\subsection{Absence of superconductivity}
\label{sec.model.nosc}

\begin{figure}[!t]
\begin{center}
\includegraphics[width=0.6\linewidth]{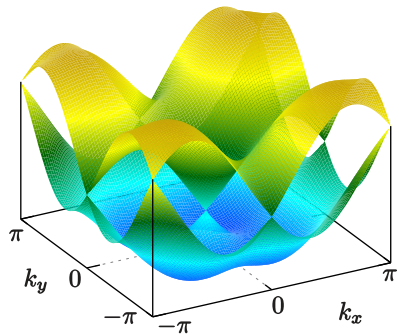}
\end{center}
\caption{
The lower and upper Rashba bands for $V_{SO} = 2 t$ in the absence of the next-nearest-neighbor hopping ($t' = 0$) and superconductivity ($\Delta_0=0$) in the first Brillouin zone for a system on the square lattice.
\label{fig.rashbaband}
}
\end{figure}

When there is no superconductivity in the system, i.e.,  if $\Delta_{0} = 0$ is assumed, from the diagonalization of the Hamiltonian $\hat{H}_{0} + \hat{H}_{SO}$, we retrieve two bands, namely, the upper and lower Rashba bands.
The eigenproblem $ (\hat{H}_{0} + \hat{H}_{SO} ) | \Psi_{{\bm k},\pm} \rangle = \varepsilon_{{\bm k},\pm} | \Psi_{{\bm k},\pm} \rangle$
gives the following eigenvalues:
\begin{eqnarray}
\label{eq.eigen}
\varepsilon_{{\bm k},\pm} = \bar{E}_{{\bm k}\alpha} \pm 2 V_{SO} \sqrt{ \sin^{2} ( k_{x} ) + \sin^{2} ( k_{y} ) },
\end{eqnarray}
and eigenstates: $| \Psi_{{\bm k},\alpha} \rangle=\hat{\Psi}_{{\bm k},\alpha}^\dag | 0 \rangle $ ($\alpha=+,-$), where
\begin{eqnarray}
\label{eq.nonint.eigenvec}
\left(
\begin{array}{c}
\hat{\Psi}_{{\bm k},+}^\dag  \\
\hat{\Psi}_{{\bm k},-}^\dag
\end{array}
\right) &=&
\frac{1}{\sqrt{ 1 + \zeta_{\bm k}^{2}}}
\left(
\begin{array}{cc}
1 & \zeta_{\bm k} \\
- \zeta_{\bm k} & 1
\end{array}
\right)
\left(
\begin{array}{c}
\hat{c}^{\dag}_{{\bm k}\uparrow} \\
\hat{c}^{\dag}_{{\bm k}\downarrow}
\end{array}
\right)
\end{eqnarray}
and
\begin{eqnarray}
\label{eq.zeta} && \zeta_{\bm k} = \\
\nonumber && \frac{ \left(\check{V}_{\bm k}\right)_{\uparrow\downarrow} }{ \dfrac{1}{2} \left( \bar{E}_{{\bm k}\uparrow} + \bar{E}_{{\bm k}\downarrow} \right) + \sqrt{ \dfrac{1}{4} \left( \bar{E}_{{\bm k}\uparrow} + \bar{E}_{{\bm k}\downarrow} \right)^{2} + \Big| \left(\check{V}_{\bm k}\right)_{\uparrow\downarrow} \Big|^{2} } } .
\end{eqnarray}
We find fourfold-degenerate minimal energy at the non-analytical points $k_{x} = \pm k_{y}$ given by the equation $\sqrt{2} \tan ( k_{x(y)} ) \left( t + 2 t' \cos ( k_{x(y)} ) \right) = 0$.
There are also four saddle points near the energy minimum points, which are located at ${\bm k} = ( 0, \pm \mbox{atan} ( V_{SO} / t ) )$ and $( \pm \mbox{atan} ( V_{SO} / t ), 0 )$.
The lower and upper Rashba bands correspond to minus ``$-$'' and plus ``$+$'' signs in above formulas, respectively,  for $V_{SO} = 2 t$ and $t' = 0$ are shown in Fig.~\ref{fig.rashbaband}.

\subsection{Including superconductivity}

Introducing the Nambu spinors $\hat{\Phi}_{\bm k} = ( \hat{c}_{{\bm k}\uparrow}, \hat{c}_{{\bm k}\downarrow}  ,  \hat{c}_{-{\bm k}\uparrow}^{\dagger} , \hat{c}_{-{\bm k}\downarrow}^{\dagger} )^{T}$, the total Hamiltonian (\ref{eq.hammomspace}) including all terms  can be rewritten in the matrix form,
\begin{eqnarray}
\hat{H} = \frac{1}{2} \sum_{\bm k} \hat{\Phi}_{\bm k}^{\dagger} \mathbb{H}_{\bm k} \hat{\Phi}_{\bm k} + \frac{1}{2} \sum_{{\bm k}\sigma} \left( \bar{E}_{{\bm k}\sigma} - U | \Delta_{0} |^{2} \gamma^{2} ( {\bm k} ) \right) ,
\label{eq.ham.mf}
\end{eqnarray}
where
\begin{eqnarray}
\label{eq.ham.matrix} && \mathbb{H}_{\bm k} = \\
\nonumber && \left(
\begin{array}{cccc}
\bar{E}_{{\bm k}\uparrow} & \left(\check{V}_{\bm k}\right)_{\uparrow\downarrow} & U\Delta_{0}\gamma( {\bm k} ) & 0 \\
\left(\check{V}_{\bm k}\right)_{\uparrow\downarrow}^{\ast} & \bar{E}_{{\bm k}\downarrow} & 0 & U\Delta_{0}\gamma( {\bm k} ) \\
U\Delta_{0}^{\ast}\gamma( {\bm k} ) & 0 & -\bar{E}_{-{\bm k}\uparrow} & -\left(\check{V}_{-{\bm k}}\right)_{\downarrow\uparrow}  \\
 0 & U\Delta_{0}^{\ast}\gamma( {\bm k} ) & -\left(\check{V}_{-{\bm k}}\right)_{\downarrow\uparrow}^{\ast} & -\bar{E}_{-{\bm k}\downarrow}
\end{array}
\right) .
\end{eqnarray}
In the above expression $\left(\check{V}_{\bm k}\right)_{\alpha\beta}$ correspond to the  matrix elements of the spin-orbit coupling matrix $ \check{V}_{\bm k}$ defined previously.
The grand canonical potential of the system is determined  by
\begin{eqnarray}
\Omega = &-& \frac{1}{2} k_{B} T \sum_{{\bm k},n=1}^{4} \ln \left( 1 + \exp \left( \frac{-\lambda_{{\bm k}n}}{k_{B}T} \right) \right) \\
\nonumber &+& \frac{1}{2} \sum_{{\bm k}\sigma} \left( \bar{E}_{{\bm k}\sigma} - U_{0} | \Delta_{0} |^{2} \gamma^{2} ( {\bm k} ) \right) .
\end{eqnarray}
where $\lambda_{{\bm k}n}$ ($n = 1,...,4$) are the eigenvalues of $\mathbb{H}_{\bm k}$ matrix, which is given by (\ref{eq.ham.matrix}).
$T$ is the absolute temperature.

\section{Numerical results}
\label{sec.numerical}

The calculations are carried out for a two dimensional square lattice of a size $N_{X} \times N_{Y} = 200 \times 200$ with the periodic boundary conditions.
In the case of the BCS state this size corresponds to the thermodynamic limit~\cite{ptok.crivelli.17}.
We find the ground state of the system as the global minimum of $\Omega$ with respect to $\Delta_0$ for a given set of model parameters $\{\mu, V_{SO}, T\}$, using the procedure described in Ref.~\cite{januszewski.ptok.15}.
We take the NN hopping ($|t|=1$) as the energy unit, while the NNN hopping is set as $t' = -0.1 t$.
As a consequence, the half-filling (i.e., $n=1$) is attained for $\mu/t = 0$ and $\mu/t =-0.4$ in the absence and in the presence of the NNN hopping, respectively.
Moreover, in all of our calculations presented here, we set the interaction strengths to $U/t = -2$ and $U/t = -1.25$ for {\it s-wave} and {\it d-wave} symmetry, respectively.
These choices of $U$ interactions give approximately equal critical temperatures $T_{c}$ at half-filling in the absence of the NNN hopping for both symmetries considered.
Nevertheless, the choice of specific values of attractive $U<0$, at least for not very large values of $|U|/t$ \cite{micnas.ranninger.90}, should not change qualitatively the results presented further in the paper.

\begin{figure}[!b]
\begin{center}
\includegraphics[width=\linewidth,height=6cm,keepaspectratio]{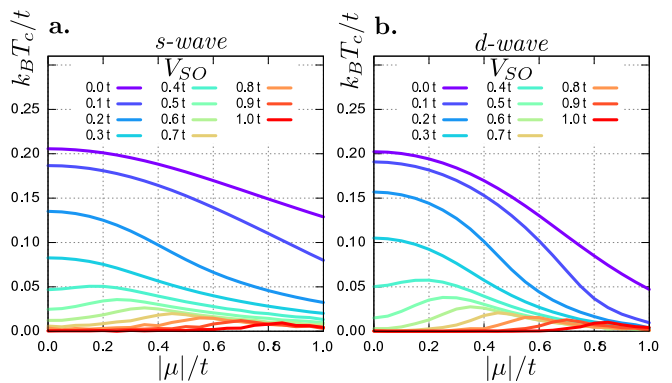}
\end{center}
\caption{
Critical temperature as a function of the chemical potential $\mu$ for several spin-orbit coupling strengths $V_{SO}$ (as labeled) for a square lattice in the absence of the next-nearest-neighbor hopping ($t' = 0$).
Two types of superconductivity  are considered: (a) {\it s-wave} and (b) {\it d-wave}.
\label{fig.ktmu}
}
\end{figure}

\subsection{Absence of the NNN hopping ($t'=0$)}

We start analyzing the influence of the spin-orbit coupling $V_{SO}$ on the critical temperature $T_{c}$ as a function of the chemical potential $\mu$ in the absence of the NNN hopping ($t' = 0$).
The electron concentration $n$  (the filling of the system) is a monotonously increasing function of $\mu$.
We define $T_{c}$ as the temperature, at which the amplitude of the superconducting order parameter $\Delta_{0}$, vanishes for a given $\{\mu,V_{SO}\}$.
As we will show $T_c=T_c(\mu,V_{SO})$ is not a trivial monotonic function of $\mu$ and $V_{SO}$.
All transitions found in the system are second-order (continuous) ones.

The phase diagrams for a few values of $V_{SO}$ are presented in Fig.~\ref{fig.ktmu}.
For $V_{SO} = 0$, we find monotonic decreasing behavior of $T_{c}$ as a function of $|\mu|$ with maximum at the half-filling (for $\mu=0$), i.e., $T_c^{max}(V_{SO}=0)=T_c(\mu=0,V_{SO}=0)$. Switching the SO coupling on, the situation changes and the maximal $T_{c}$ is located away from the half-filling (for $\mu \neq 0$), i.e., $T_c^{max}(V_{SO}\neq0)=T_c(\mu\neq0,V_{SO}\neq0)$.
For fixed $V_{SO}\neq0$, $T_c$ is a non-monotonous function of $\mu$ with a local minimum at $\mu=0$.
However, for small enough SO values, smaller than $V_{SO} \approx 0.36 t$, the behavior changes very slightly and it cannot be seen unambiguously.
The continuous increase of $T_{c}$ as the doping increases leads to a growth of $T_{c}$ due to the presence of the SO coupling.
The same qualitative behavior is observed in both symmetry cases, but there are some quantitative differences between them.
Particularly, from Fig.~\ref{fig.ktmu} it is seen that the superconductivity is more sensitive to the doping for {\it d-wave} symmetry, i.e., for fixed $V_{SO}$ temperature $T_{c}$ varies more with changing $\mu$ in the presence of {\it d-wave} pairing.
In contrary, $V_{SO}$ suppresses {\it s-wave} superconductivity stronger (i.e., $T_{c}(\mu=0,V_{SO})$ at half-filling decreases faster with increasing of $V_{SO}$ for the {\it s-wave} case).

\begin{figure}[!t]
\begin{center}
\includegraphics[width=\linewidth,height=6cm,keepaspectratio]{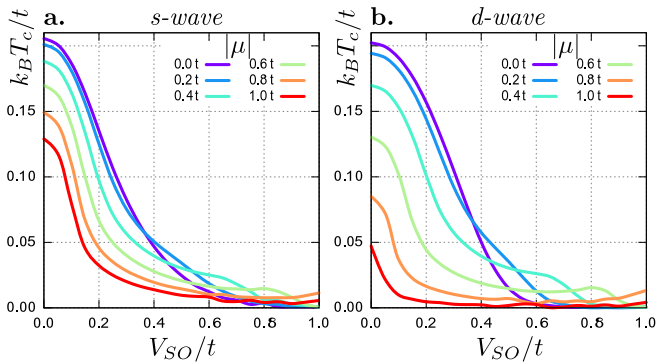}
\end{center}
\caption{
Critical temperature as a function of the spin-orbit coupling $V_{SO}$ for several chemical potential values $\mu$ (as labeled) in the absence of the next-nearest-neighbor hopping ($t' = 0$).
Two types of superconductivity  are considered: (a) {\it s-wave} and (b) {\it d-wave}.
\label{fig.ktvso}
}
\end{figure}

\begin{figure}[!b]
\begin{center}
\includegraphics[width=\linewidth,height=6cm,keepaspectratio]{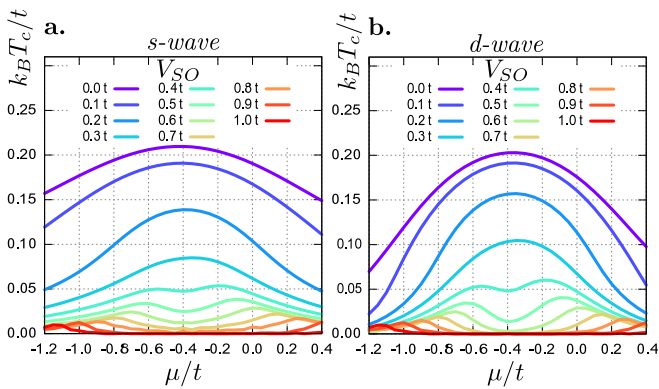}
\end{center}
\caption{
Critical temperature as a function of the chemical potential $\mu$ for several spin-orbit coupling values $V_{SO}$ (as labeled) for a square lattice in the presence of the next-nearest-neighbor hopping term ($t' = - 0.1 t$).
Two types of superconductivity are considered: (a) {\it s-wave} and (b) {\it d-wave}.
The half-filling condition corresponds to $\mu=-0.4 t$.
\label{fig.ktmunnn}
}
\end{figure}

In order to gain a better understanding of the situation, we plot $T_{c}$, but this time as a function of the SO coupling for several values of the chemical potential, as shown in Fig.~\ref{fig.ktvso}.
It is found that for fixed $\mu$ the $T_c$ is not a monotonously decreasing function of $V_{SO}$ (if the chemical potential, or equivalently doping, is away from half-filing).
For small values of $V_{SO}$, an increase of the SO coupling reduces $T_c$.
However, at sufficiently large $\mu$ and for larger values of the SO coupling, the temperature $T_{c}$ increases again.
This revival is clearly seen in this {\it cross-section} of the $k_BT$ versus $V_{SO}$ phase diagrams,
where the critical temperature for larger $|\mu|$ exceeds the values obtained at lower values of $\mu$, particularly those derived for the half-filling.

\begin{figure}[!t]
\begin{center}
\includegraphics[width=\linewidth,height=12cm,keepaspectratio]{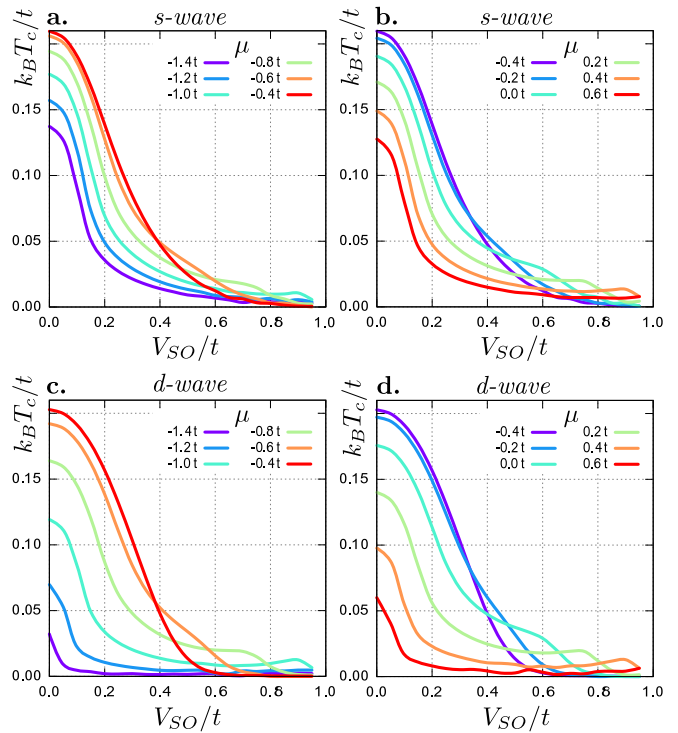}
\end{center}
\caption{
Critical temperature as a function of the spin-orbit coupling, $V_{SO}$, for several chemical potential values, $\mu$, (as labeled) in the presence of the NNN hopping ($t' = - 0.1 t$).
Two types of superconductivity  are considered: (a), (b) {\it s-wave} and (c), (d) {\it d-wave}.
The half-filling condition corresponds to $\mu=-0.4t$.
Panels (a) and (c) presents curves for underdoped system, whereas panels (b) and (d) are obtained in overdoped regime.
\label{fig.ktvsonnn}
}
\end{figure}

\subsection{Presence of the NNN hopping ($t'\neq0$)}

Next, we investigate the behavior of the system in the presence of the NNN hopping ($t'=-0.1t$).
In such a case the dependence of $T_c$ a function of $\mu$ losses its symmetry around half-filling, see Fig.~\ref{fig.ktmunnn} and Fig.~\ref{fig.ktvsonnn}.
Similarly as before, we observe a substantial influence of the spin-orbit coupling on $\mu$-dependence of $T_{c}$, which is stronger in the {\it d-wave} symmetry case.
Again, we obtain a growth of $T_{c}$ over its half-filling value as the chemical potential changes away from the half-filling for  fixed $V_{SO}>0$ (Fig.~\ref{fig.ktmunnn}).
However, the value of maximal $T_c$ in the overdoped regime (i.e., $\mu>-0.4t$) is larger than that obtained for underdoped system (i.e., $\mu<-0.4t$).
$T_c$ as a function of $V_{SO}$ for fixed $\mu$ exhibits similar properties as discussed for $t'=0$ previously, although the values of $T_c$ for underdoped and overdoped system differ from each other (Fig.~\ref{fig.ktvsonnn})
As we indicated before, the half-filling condition $n=1$ for the model parametr used corresponds to $\mu=-0.4t$.
One should remember that the value of $T_{c}$, for a given $\{\mu,V_{SO}\}$, depends also on $t'$.

\section{Discussion}
\label{sec.discuss}

Now we will discuss our results in terms of {\it (i)} the density of states of the non-interacting system and {\it (ii)} the  non-trivial superconductivity induced by the SO coupling.
In relation to superconducting state, the density of states of the non-interacting system at the Fermi level  affects the  critical temperatures and critical magnetic fields.
This relation between these macroscopic and microscopic quantities has been described in the pioneering papers of Bardeen, Cooper, and Schrieffer~\cite{bardeen.cooper.57a,bardeen.cooper.57b}.
On the other hand, the existence of the SO coupling in the system leads to a mixing of electron states with opposite spins, what is clearly seen in the expressions for the eigenvectors of non-interacting system given by Eq.~(\ref{eq.nonint.eigenvec}).
As a consequence, we can expect a realisation of the non-trivial triplet pairing in the system~\cite{sato.takahasi.10,
gorkov.rashba.01,zhang.tewari.08,alicae.10,seo.han.12,yu.wu.16}.

\subsection{Density of state}

To understand the behavior of the system described in the previous section, i.e., the dependence of $T_c$ as a function of $\mu$ and $V_{SO}$, we calculate the non-interacting partial density of states (DOS)~\cite{maska.93}:
\begin{eqnarray}
\rho_{\pm} ( E ) = \frac{1}{N} \sum_{\bm k} \delta \left( \varepsilon_{{\bm k},\pm} - E \right) ,
\end{eqnarray}
where the eigenvalues $\varepsilon_{{\bm k},\pm}$ of the $\hat{H}_{0} + \hat{H}_{SO}$ are given by Eq.~(\ref{eq.eigen}).
In our case the total density of states is given by $\rho(E) = \rho_{+}(E) + \rho_{-}(E)$.

\begin{figure}[!b]
\begin{center}
\includegraphics[width=\linewidth,height=6cm,keepaspectratio]{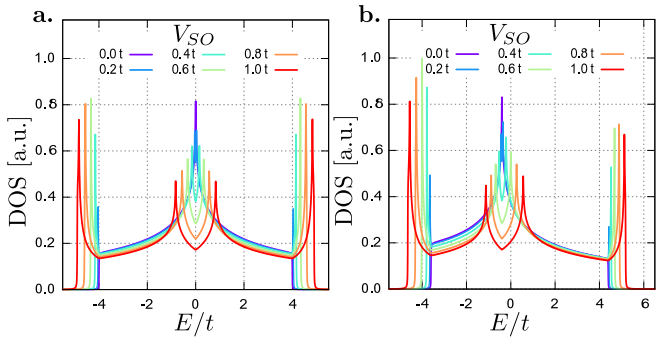}
\end{center}
\caption{
The non-interacting total density of states for (a) $t'=0$ and (b) $t'= - 0.1t$ and for several values of $V_{SO}$ (as labeled).
The Fermi level is located at $E=-4t'$ for the half-filling.
The shape of the non-interacting density of states is not dependent on $\mu$.
Only the Fermi level $E_F$ depends on $\mu$ ($E_F =\mu-4t'$).
\label{fig.dos}
}
\end{figure}

\begin{figure}[!t]
\begin{center}
\includegraphics[width=\linewidth,height=10cm,keepaspectratio]{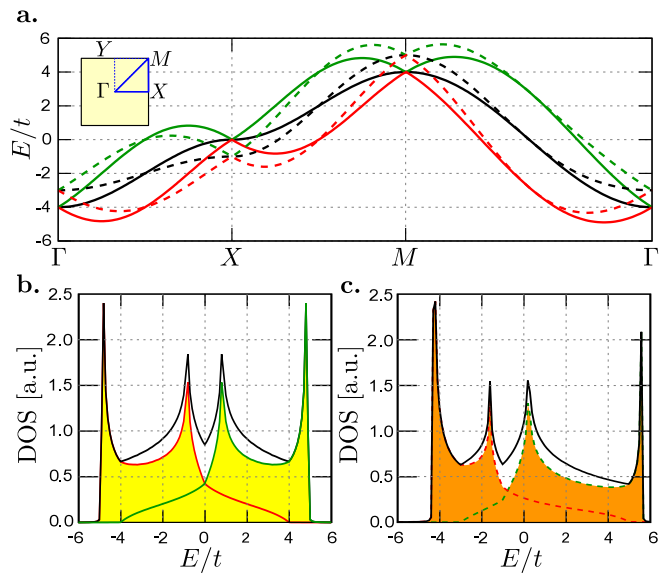}
\end{center}
\caption{
(a) The band structure along high symmetry points $\Gamma$--X--M--$\Gamma$ (left inset for locations of the points) for $t'=0$ (solid lines) and $t'= - 0.25t$ (dashed lines):  the electron band structure (for $V_{SO}=0$) is represented by the black lines, whereas the lower and the upper Rashba bands for $V_{SO} = 1 t$ are indicated by red and green lines, respectively.
The lower panels show the total ($\rho(E)$, black line) and partial ($\rho_{\mp}(E)$, red and green lines for the lower and the upper Rashba bands) densities  of states for (b) $t'=0$ and  (c) $t'=-0.25t$, with spin-orbit coupling $V_{SO} = 1 t$. 
\label{fig.bands}
}
\end{figure}

In Fig.~\ref{fig.dos}, we present DOS for both $t'=0$ and $t'\neq0$ cases for different values of $V_{SO}$.
We retrieve symmetric profiles around $E=0$ in the absence of the NNN term as expected (Fig.~\ref{fig.dos}(a)).
On the contrary, in the presence of the NNN hopping, the symmetry around $E=-4t'$ is lost.
This fact reflects the behavior already retrieved in the $T$ versus $\mu$ phase diagram presented in Figs.~\ref{fig.ktmu} and~\ref{fig.ktmunnn}.
From the results presented in Fig.~\ref{fig.dos} one can conclude that a non-zero SO coupling $V_{SO}$ leads to
{\it (i)} a division of the van Hove (central) peak in $\rho(E)$ at $E=-4t'$, and
{\it (ii)} the emergence of additional peaks near both band edges.
The changes reported in the DOS are the result of the modifications of the band structure due to the SO coupling (Fig.~\ref{fig.bands}(a)).
The double-peak structure of $\rho(E)$ near $E\approx-4t'$ is the result of an existence of energy local minima (maxima) of the dispersion relation near points X and Y of the  Brillouin zone in the lower (upper) Rashba band (cf. also Fig.~\ref{fig.rashbaband} for $t'=0$).
On the contrary, the peaks at the edges are due to the existence of four saddle points in every Rashba band with energies
$E_{\mbox{sad}} = \mp 2 t \left( 1 + \sqrt{ 1 + (V_{SO}/t)^{2} } \right)$ near $\Gamma$ and M points, where $\mp$ signs correspond to the lower and upper Rashba bands, respectively.

Assuming that the critical temperature is given by the standard BCS formula, then one gets $T_{c} \sim \exp ( - 1 / \rho ( E_{F} ) )$~\cite{bardeen.cooper.57a,bardeen.cooper.57b}.
It is rather justified for the values of $U/t$ considered in the present work.
Using the relation between total and partial DOS, we obtain that $T_{c} \sim \exp [ - 1 / ( \rho_{+} ( E_{F} ) + \rho_{-} ( E_{F} ) ) ]$,
where $E_{F} = \mu - 4 t' $ in a case of the square lattice with the NNN hopping considered here.
To illustrate this fact, the two partial DOS are presented in Figs.~\ref{fig.bands}(b) and (c).
A direct comparison with Fig.~\ref{fig.dos} shows that the greatest influence on the total DOS comes only from one Rashba band.
In the case of the underdoped system it is the lower Rashba band, while in the overdoped case the upper Rashba band is the crucial one.

In the  results presented in this paper, for $T_{c}$ as a function of $\mu$ and for $V_{SO}\neq 0$, we have not found any consequences of the existence of the additional narrow peaks in the DOS $\rho(E)$  at both edges of the Rashba bands.
It can be due to the fact that these peaks are located solely at the Fermi level for very small either electron or hole concentrations in the system (i.e., at $n\approx0$ or $n\approx2$, respectively).
However, this behavior can have an important role in extremally dilute systems, e.g., in the BCS-BEC crossover region~\cite{chen.gong.12,shi.rosenberg.16,lee.kim.17}.

\subsection{Non-trivial pairing}

As it was indicated above, in the presence of the SO coupling the electron spin is not longer a good quantum number.
Despite the fact that we have labeled the system under our consideration to present either {\it s-wave} or  {\it d-wave} symmetry of the superconducting order parameter, the SO coupling can {\it effectively introduce} a {\it p-wave} superconductivity.
This possibility is well known and described in the literature, e.g., Refs.~\cite{sato.takahasi.10,gorkov.rashba.01,zhang.tewari.08,alicae.10,seo.han.12,yu.wu.16}.
In this aspect a mutual relation between the pairing in electron and quasiparticle spaces is important (see Appendix~\ref{app.pwave}).
As a consequence of this matter, it is possible to realise the non-trivial {\it triplet} pairing in the considered system in the presence of the SO coupling, even if conventional pairing (for both initial gap symmetries, i.e. {\it s-wave} and {\it d-wave}) is a source of superconductivity.

This issue can be described by a transformation, which changes the basis from the original one into the helicity basis and one can find a ratio (cf. Eq.~(\ref{eq.trip.sing.ratio})) between triplet and singlet pairing (more details can be found in Appendix~\ref{app.pwave}).
This ratio is a non-trivial function of the doping $\mu$, the NNN hopping $t'$,  and the SO coupling $V_{SO}$.
We found that this ratio is an increasing function of $V_{SO}$ and doping in the neighborhood of the half-filing.
It changes from zero (at half-filing and small $V_{SO}$) to a few tens (at optimal doping, i.e., doping for which $T_c$ is the largest one at fixed $V_{SO}$).
Moreover, it depends on the momentum as consequence of non-isotropic symmetry of the gap in considered system (i.e., for {\it d-wave} case).

\section{Summary and final remarks}
\label{sec.summary}

The spin-orbit coupling can lead to different unexpected behaviors in various systems.
In this paper, we discuss the influence of the spin-orbit coupling on superconducting states in the presence of on-site ({\it s-wave}) or inter-site ({\it d-wave}) pairing in a monolayer with spin-orbit coupling induced by the proximity effects.
In particular, we discuss the temperature versus doping phase diagram in detail.
In our work, tuning the doping away from the half-filling, we demonstrate that the critical temperature can be a non-monotonic function of the spin-orbit coupling.
Moreover, we have shown that for fixed value of the spin-orbit interaction the maximal value of the critical temperature is obtained for the underdoped or overdoped regimes, i.e., away from half-filling.
Therefore, our results highlight the effects of the spin-orbit coupling on superconducting properties of the system.

These results may be particularly relevant due to the feasibility of experimental realization of superconducting nano- and spintronics devices where a temperature dependence on the spin-orbit coupling using doped systems can enhance critical temperatures $T_{c}$ significantly rather than those at half-filling.
It could be of a great importance due to the fact that for applications of superconducting materials systems with larger $T_{c}$ are preferred.

The modifications of the $T_{c}$ by the spin-orbit coupling have been also discussed within the Werthamer-Helfand-Hohenberg (WHH) theory~\cite{whh}, which describes the orbitally limited upper critical field of dirty II-type superconductors.
In the WHH theory $T_{c}$ is found as a function of a parameter describing the system~\cite{lei.hu.10}, e.g., it can be the SO coupling.
Similarly as the findings of the present work, the WHH theory predicts that an increase of the spin-orbit coupling can lead to an increase of $T_{c}$~\cite{wolfffabris.lei.14}.
Moreover, similar behavior of $T_{c}$ can be observed in systems with non-trivial {\it p-wave} pairing~\cite{weng.hu.16}.

Presented results have been described in the context of solid state physics, e.g., a superconducting monolayer on a surface of an insulator with strong spin-orbit coupling (e.g., topological insulator).
In this system, as a consequence of the proximity effects, the SO is induced in the layer and affects the physical properties of the superconductor.
However, the investigations of the effects of the SO coupling on superconductivity beyond the described heterostructure can be performed with ultracold atomic gases on the optical lattices.~\cite{bloch.dalibard.08}.
The realisation of the artificial SO coupling in such systems is possible~\cite{zhang.tewari.08,jiang.liu.11,galitski.spielman.13,fu.huang.14,grusdt.li.17}.
Such experiments are important towards to an experimental realization of atomic superfluids with topological excitations.

\begin{acknowledgments}
The authors are thankful to T. Doma\'{n}ski, Sz. G\l{}odzik, A. Kobia\l{}ka, P. Piekarz, and K. I. Wysoki\'{n}ski for very fruitful discussions and comments.
K.R. acknowledges the support from CIBioFi and the Colombian Science, Technology and Innovation Fundation --- COLCIENCIAS ``Francisco Jos\'e de Caldas'' under project 1106-712-49884 (contract No.264-2016) and --- General Royalties System (Fondo CTeI-SGR) under contract No. BPIN 2013000100007.
The support from UMO-2016/20/S/ST3/00274 (A.P.), UMO-2016/21/D/ST3/03385 (K.J.K.) and
UMO-2017/24/C/ST3/00276 (K.J.K.) projects by Narodowe Centrum Nauki (NCN, National Science Centre, Poland) is also acknowledged.
\end{acknowledgments}

\appendix

\section{Non-trivial superconductivity induced by the spin-orbit coupling}

\label{app.pwave}

\begin{figure}[!t]
\begin{center}
\includegraphics[width=\linewidth]{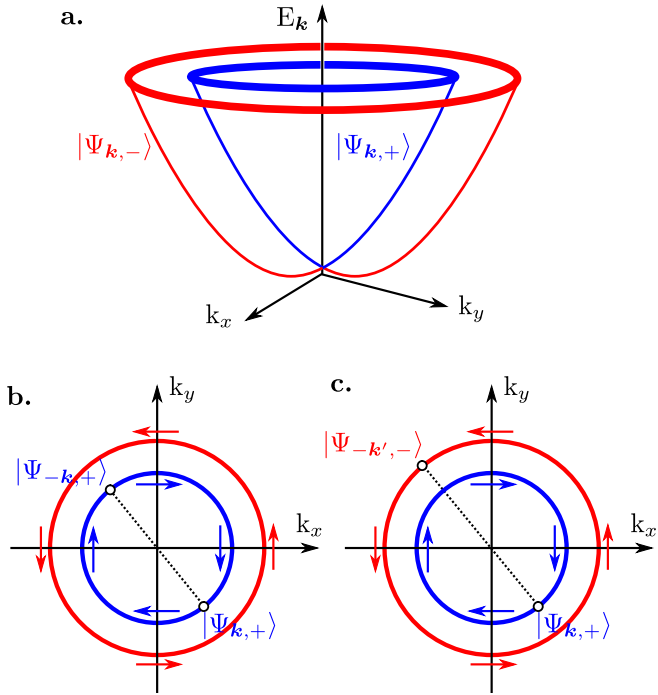}
\end{center}
\caption{Schematic illustration of the influence of the SO coupling on the band structure.
(a) In the presence of the SO the spin degeneracy is lifted  and two Rashba bands  $| \Psi_{{\bm k},\alpha} \rangle$ with different indexes ($\alpha=-,+$) of pseudospin (helical spin texture) arise (the arrows mark the pseudospin directions).
The lower $| \Psi_{{\bm k},-} \rangle$ and upper $| \Psi_{{\bm k},+} \rangle$ Rashba bands are denoted by red and blue lines, respectively.
Examples of the intra-band (panel (b)) and inter-band (panel (c)) pairing are shown.
\label{fig.schem2}
}
\end{figure}

Formally, in the original basis, the singlet Cooper pairs are described by the Hamiltonian~(\ref{eq.hammomspace}).
It formally corresponds to a pairing of electrons with opposite spin and momentum.
An existence of the SO coupling in the system leads to a mixing of the spins, what is clearly shown by Eq.~(\ref{eq.nonint.eigenvec}).
As a consequence,  in the non-interacting system,  one can discuss an existence of the lower and upper Rashba bands, what have been described in Sec.~\ref{sec.model.nosc}.

Technically, the Hamiltonian of the system in the presence of superconductivity can be rewritten in a helicity basis of  eigenstates~(\ref{eq.nonint.eigenvec}).
Then, the matrix form~(\ref{eq.ham.matrix}) of Hamiltonian is formally given as $\hat{H} = \frac{1}{2} \sum_{\bm k} \hat{\varphi}_{\bm k}^{\dagger} \bar{\mathbb{H}}_{\bm k} \hat{\varphi}_{\bm k} + \mbox{const.}$, where
\begin{eqnarray}
\bar{\mathbb{H}}_{\bm k} = \left(
\begin{array}{cccc}
\varepsilon_{{\bm k},+} & 0 & \tilde{\Delta}_{++} ( {\bm k} ) & \tilde{\Delta}_{+-} ( {\bm k} ) \\
0 & \varepsilon_{{\bm k},-} & \tilde{\Delta}_{-+} ( {\bm k} ) & \tilde{\Delta}_{--} ( {\bm k} ) \\
\tilde{\Delta}_{++}^{\ast} ( {\bm k} ) & \tilde{\Delta}_{-+}^{\ast} ( {\bm k} ) & -\varepsilon_{-{\bm k},+} & 0 \\
\tilde{\Delta}_{+-}^{\ast} ( {\bm k} ) & \tilde{\Delta}_{--}^{\ast} ( {\bm k} ) & 0 & -\varepsilon_{-{\bm k},-}
\end{array}
\right),
\end{eqnarray}
$\hat{\varphi}_{\bm k} = ( \hat{\Psi}_{{\bm k},+}, \hat{\Psi}_{{\bm k},-} , \hat{\Psi}_{-{\bm k},+}^{\dagger} , \hat{\Psi}_{-{\bm k},-}^{\dagger} )^{T}$, and $\hat{\Psi}_{{\bm k},\pm}$ are defined by (\ref{eq.nonint.eigenvec}), for details cf. Ref.~\cite{seo.han.12}.
Similarly as previously, ''plus'' or ''minus'' signs correspond to  helicity basis (i.e., the Rashba bands).
Superconducting order parameters (SOP) $\tilde{\Delta}_{++}$ and $\tilde{\Delta}_{--}$ describes the intraband pairing, whereas parameters $\tilde{\Delta}_{+-}$ and $\tilde{\Delta}_{-+}$ are associated with the interband pairing (what is schematically shown in Fig.~\ref{fig.schem2}).
Notice that they are dependent on ${\bm k}$.
The new SOP in the helicity basis are defined as $\tilde{\Delta}_{\alpha\beta}=\langle \hat{\Psi}^\dag_{-{\bm k},\alpha} \hat{\Psi}_{{\bm k},\beta} \rangle$.
The first type of pairing ($\alpha=\beta$) corresponds to singlet pairing of electrons in the helicity
basis (Fig.~\ref{fig.schem2}(b), it contains not only {\it s-wave} but also {\it d-wave} and higher angular momentum contributions), while the second one ($\alpha\neq\beta$) is associated with the triplet pairing in the helicity
basis (Fig.~\ref{fig.schem2}(c), it contains not only {\it p-wave} but also {\it f-wave} and higher angular momentum contributions).

Using transformation~(\ref{eq.nonint.eigenvec}), one can express the SOPs in the helicity
basis by the SOP in the original basis~\cite{linder.sudbo.09,ptok.14}:
\begin{eqnarray}
\nonumber \tilde{\Delta}_{++} ( {\bm k} ) &=& \tilde{\Delta}_{--} ( {\bm k} ) = \\
&=& U \Delta_{0} \gamma ( {\bm k} ) \frac{ 1 + \zeta_{\bm k} \zeta_{-{\bm k}} }{ \sqrt{1+\zeta_{\bm k}^{2}} \sqrt{1+\zeta_{-{\bm k}}^{2}} } , \\
\nonumber \tilde{\Delta}_{+-} ( {\bm k} ) &=& - \tilde{\Delta}_{+-} ( {\bm k} ) = \\
&=& U \Delta_{0} \gamma ( {\bm k} ) \frac{ \zeta_{\bm k} - \zeta_{-{\bm k}} }{ \sqrt{1+\zeta_{\bm k}^{2}} \sqrt{1+\zeta_{-{\bm k}}^{2}} } ,
\end{eqnarray}
where $\zeta_{\bm k}$ is given by Eq.~(\ref{eq.zeta})
From this relations, the ratio between triplet and singlet component of the non-isotropic gap superconductor depends on momentum and can be express as:
\begin{eqnarray}
\label{eq.trip.sing.ratio} \eta_{\bm k} = \frac{\tilde{\Delta}_{+-} ( {\bm k} )}{ \tilde{\Delta}_{++} ( {\bm k} ) } = \frac{ \zeta_{\bm k} - \zeta_{-{\bm k}} }{ 1 + \zeta_{\bm k} \zeta_{-{\bm k}} } .
\end{eqnarray}
Notice also the fact that $|\tilde{\Delta}_{+-} ( {\bm k} )|^2 +  |\tilde{\Delta}_{++} ( {\bm k} )|^2 = |\Delta_0|^2  $.

\bibliography{biblio}

\begin{thebibliography}{93}%
\makeatletter
\providecommand \@ifxundefined [1]{%
 \@ifx{#1\undefined}
}%
\providecommand \@ifnum [1]{%
 \ifnum #1\expandafter \@firstoftwo
 \else \expandafter \@secondoftwo
 \fi
}%
\providecommand \@ifx [1]{%
 \ifx #1\expandafter \@firstoftwo
 \else \expandafter \@secondoftwo
 \fi
}%
\providecommand \natexlab [1]{#1}%
\providecommand \enquote  [1]{``#1''}%
\providecommand \bibnamefont  [1]{#1}%
\providecommand \bibfnamefont [1]{#1}%
\providecommand \citenamefont [1]{#1}%
\providecommand \href@noop [0]{\@secondoftwo}%
\providecommand \href [0]{\begingroup \@sanitize@url \@href}%
\providecommand \@href[1]{\@@startlink{#1}\@@href}%
\providecommand \@@href[1]{\endgroup#1\@@endlink}%
\providecommand \@sanitize@url [0]{\catcode `\\12\catcode `\$12\catcode
  `\&12\catcode `\#12\catcode `\^12\catcode `\_12\catcode `\%12\relax}%
\providecommand \@@startlink[1]{}%
\providecommand \@@endlink[0]{}%
\providecommand \url  [0]{\begingroup\@sanitize@url \@url }%
\providecommand \@url [1]{\endgroup\@href {#1}{\urlprefix }}%
\providecommand \urlprefix  [0]{URL }%
\providecommand \Eprint [0]{\href }%
\providecommand \doibase [0]{http://dx.doi.org/}%
\providecommand \selectlanguage [0]{\@gobble}%
\providecommand \bibinfo  [0]{\@secondoftwo}%
\providecommand \bibfield  [0]{\@secondoftwo}%
\providecommand \translation [1]{[#1]}%
\providecommand \BibitemOpen [0]{}%
\providecommand \bibitemStop [0]{}%
\providecommand \bibitemNoStop [0]{.\EOS\space}%
\providecommand \EOS [0]{\spacefactor3000\relax}%
\providecommand \BibitemShut  [1]{\csname bibitem#1\endcsname}%
\let\auto@bib@innerbib\@empty
\bibitem [{\citenamefont {de~Gennes}(1964)}]{degennes.64}%
  \BibitemOpen
  \bibfield  {author} {\bibinfo {author} {\bibfnamefont {P.~G.}\ \bibnamefont
  {de~Gennes}},\ }\bibfield  {title} {\enquote {\bibinfo {title} {Boundary
  effects in superconductors},}\ }\href {\doibase 10.1103/RevModPhys.36.225}
  {\bibfield  {journal} {\bibinfo  {journal} {Rev. Mod. Phys.}\ }\textbf
  {\bibinfo {volume} {36}},\ \bibinfo {pages} {225} (\bibinfo {year}
  {1964})}\BibitemShut {NoStop}%
\bibitem [{\citenamefont {Holm}\ and\ \citenamefont
  {Meissner}(1932)}]{holm.meissner.32}%
  \BibitemOpen
  \bibfield  {author} {\bibinfo {author} {\bibfnamefont {R.}~\bibnamefont
  {Holm}}\ and\ \bibinfo {author} {\bibfnamefont {W.}~\bibnamefont
  {Meissner}},\ }\bibfield  {title} {\enquote {\bibinfo {title} {Messungen mit
  {Hilfe} von fl{\"u}ssigem {Helium}. {XIII}},}\ }\href {\doibase
  10.1007/BF01340420} {\bibfield  {journal} {\bibinfo  {journal} {Z. Physik}\
  }\textbf {\bibinfo {volume} {74}},\ \bibinfo {pages} {715} (\bibinfo {year}
  {1932})}\BibitemShut {NoStop}%
\bibitem [{\citenamefont {Anderson}\ and\ \citenamefont
  {Rowell}(1963)}]{anderson.rowell.63}%
  \BibitemOpen
  \bibfield  {author} {\bibinfo {author} {\bibfnamefont {P.~W.}\ \bibnamefont
  {Anderson}}\ and\ \bibinfo {author} {\bibfnamefont {J.~M.}\ \bibnamefont
  {Rowell}},\ }\bibfield  {title} {\enquote {\bibinfo {title} {Probable
  observation of the {Josephson} superconducting tunneling effect},}\ }\href
  {\doibase 10.1103/PhysRevLett.10.230} {\bibfield  {journal} {\bibinfo
  {journal} {Phys. Rev. Lett.}\ }\textbf {\bibinfo {volume} {10}},\ \bibinfo
  {pages} {230} (\bibinfo {year} {1963})}\BibitemShut {NoStop}%
\bibitem [{\citenamefont {Josephson}(1962)}]{josephson.62}%
  \BibitemOpen
  \bibfield  {author} {\bibinfo {author} {\bibfnamefont {B.D.}\ \bibnamefont
  {Josephson}},\ }\bibfield  {title} {\enquote {\bibinfo {title} {Possible new
  effects in superconductive tunnelling},}\ }\href {\doibase
  10.1016/0031-9163(62)91369-0} {\bibfield  {journal} {\bibinfo  {journal}
  {Phys. Lett.}\ }\textbf {\bibinfo {volume} {1}},\ \bibinfo {pages} {251}
  (\bibinfo {year} {1962})}\BibitemShut {NoStop}%
\bibitem [{\citenamefont {\v{Z}uti\'{c}}\ \emph {et~al.}(2004)\citenamefont
  {\v{Z}uti\'{c}}, \citenamefont {Fabian},\ and\ \citenamefont
  {Das~Sarma}}]{zutic.fabian.04}%
  \BibitemOpen
  \bibfield  {author} {\bibinfo {author} {\bibfnamefont {I.}~\bibnamefont
  {\v{Z}uti\'{c}}}, \bibinfo {author} {\bibfnamefont {J.}~\bibnamefont
  {Fabian}}, \ and\ \bibinfo {author} {\bibfnamefont {S.}~\bibnamefont
  {Das~Sarma}},\ }\bibfield  {title} {\enquote {\bibinfo {title} {Spintronics:
  Fundamentals and applications},}\ }\href {\doibase 10.1103/RevModPhys.76.323}
  {\bibfield  {journal} {\bibinfo  {journal} {Rev. Mod. Phys.}\ }\textbf
  {\bibinfo {volume} {76}},\ \bibinfo {pages} {323} (\bibinfo {year}
  {2004})}\BibitemShut {NoStop}%
\bibitem [{\citenamefont {Bandyopadhyay}\ and\ \citenamefont
  {Cahay}(2004)}]{bandyopadhyay.cahay.04}%
  \BibitemOpen
  \bibfield  {author} {\bibinfo {author} {\bibfnamefont {S.}~\bibnamefont
  {Bandyopadhyay}}\ and\ \bibinfo {author} {\bibfnamefont {M.}~\bibnamefont
  {Cahay}},\ }\bibfield  {title} {\enquote {\bibinfo {title} {Reexamination of
  some spintronic field-effect device concepts},}\ }\href {\doibase
  10.1063/1.1784042} {\bibfield  {journal} {\bibinfo  {journal} {Appl. Phys.
  Lett.}\ }\textbf {\bibinfo {volume} {85}},\ \bibinfo {pages} {1433} (\bibinfo
  {year} {2004})}\BibitemShut {NoStop}%
\bibitem [{\citenamefont {Naber}\ \emph {et~al.}(2007)\citenamefont {Naber},
  \citenamefont {Faez},\ and\ \citenamefont {van~der Wiel}}]{naber.feaz.07}%
  \BibitemOpen
  \bibfield  {author} {\bibinfo {author} {\bibfnamefont {W.~J.~M.}\
  \bibnamefont {Naber}}, \bibinfo {author} {\bibfnamefont {S.}~\bibnamefont
  {Faez}}, \ and\ \bibinfo {author} {\bibfnamefont {W.~G.}\ \bibnamefont
  {van~der Wiel}},\ }\bibfield  {title} {\enquote {\bibinfo {title} {Organic
  spintronics},}\ }\href {\doibase http://doi.org/10.1088/0022-3727/40/12/R01}
  {\bibfield  {journal} {\bibinfo  {journal} {J. Phys. D: Appl. Phys.}\
  }\textbf {\bibinfo {volume} {40}},\ \bibinfo {pages} {R205} (\bibinfo {year}
  {2007})}\BibitemShut {NoStop}%
\bibitem [{\citenamefont {Bandyopadhyay}\ and\ \citenamefont
  {Cahay}(2015)}]{bandyopadhyay.cahay.15}%
  \BibitemOpen
  \bibfield  {author} {\bibinfo {author} {\bibfnamefont {S.}~\bibnamefont
  {Bandyopadhyay}}\ and\ \bibinfo {author} {\bibfnamefont {M.}~\bibnamefont
  {Cahay}},\ }\href@noop {} {\emph {\bibinfo {title} {Introduction to
  spintronics}}}\ (\bibinfo  {publisher} {CRC press},\ \bibinfo {year}
  {2015})\BibitemShut {NoStop}%
\bibitem [{\citenamefont {Dresselhaus}(1955)}]{dresselhaus.55}%
  \BibitemOpen
  \bibfield  {author} {\bibinfo {author} {\bibfnamefont {G.}~\bibnamefont
  {Dresselhaus}},\ }\bibfield  {title} {\enquote {\bibinfo {title} {Spin-orbit
  coupling effects in zinc blende structures},}\ }\href {\doibase
  10.1103/PhysRev.100.580} {\bibfield  {journal} {\bibinfo  {journal} {Phys.
  Rev.}\ }\textbf {\bibinfo {volume} {100}},\ \bibinfo {pages} {580} (\bibinfo
  {year} {1955})}\BibitemShut {NoStop}%
\bibitem [{\citenamefont {Rashba}(1960)}]{rashba.60}%
  \BibitemOpen
  \bibfield  {author} {\bibinfo {author} {\bibfnamefont {E.~I.}\ \bibnamefont
  {Rashba}},\ }\bibfield  {title} {\enquote {\bibinfo {title} {Properties of
  semiconductors with an extremum loop. 1. {Cyclotron} and combinational
  resonance in a magnetic field perpendicular to the plane of the loop},}\
  }\href@noop {} {\bibfield  {journal} {\bibinfo  {journal} {Sov. Phys. Solid
  State}\ }\textbf {\bibinfo {volume} {2}},\ \bibinfo {pages} {1224} (\bibinfo
  {year} {1960})}\BibitemShut {NoStop}%
\bibitem [{\citenamefont {Bychkov}\ and\ \citenamefont
  {Rashba}(1984)}]{bychkow.rashba.84}%
  \BibitemOpen
  \bibfield  {author} {\bibinfo {author} {\bibfnamefont {Yu.~A.}\ \bibnamefont
  {Bychkov}}\ and\ \bibinfo {author} {\bibfnamefont {E.~I.}\ \bibnamefont
  {Rashba}},\ }\bibfield  {title} {\enquote {\bibinfo {title} {Properties of a
  {2D} electron gas with lifted spectral degeneracy},}\ }\href@noop {}
  {\bibfield  {journal} {\bibinfo  {journal} {JEPT Lett.}\ }\textbf {\bibinfo
  {volume} {39}},\ \bibinfo {pages} {78} (\bibinfo {year} {1984})}\BibitemShut
  {NoStop}%
\bibitem [{\citenamefont {Bansil}\ \emph {et~al.}(2016)\citenamefont {Bansil},
  \citenamefont {Lin},\ and\ \citenamefont {Das}}]{bansil.lin.16}%
  \BibitemOpen
  \bibfield  {author} {\bibinfo {author} {\bibfnamefont {A.}~\bibnamefont
  {Bansil}}, \bibinfo {author} {\bibfnamefont {H.}~\bibnamefont {Lin}}, \ and\
  \bibinfo {author} {\bibfnamefont {T.}~\bibnamefont {Das}},\ }\bibfield
  {title} {\enquote {\bibinfo {title} {{\it Colloquium}: Topological band
  theory},}\ }\href {\doibase 10.1103/RevModPhys.88.021004} {\bibfield
  {journal} {\bibinfo  {journal} {Rev. Mod. Phys.}\ }\textbf {\bibinfo {volume}
  {88}},\ \bibinfo {pages} {021004} (\bibinfo {year} {2016})}\BibitemShut
  {NoStop}%
\bibitem [{\citenamefont {Jungwirth}\ \emph {et~al.}(2012)\citenamefont
  {Jungwirth}, \citenamefont {Wunderlich},\ and\ \citenamefont
  {Olejnik}}]{jungwirth.wunderlich.12}%
  \BibitemOpen
  \bibfield  {author} {\bibinfo {author} {\bibfnamefont {T.}~\bibnamefont
  {Jungwirth}}, \bibinfo {author} {\bibfnamefont {J.}~\bibnamefont
  {Wunderlich}}, \ and\ \bibinfo {author} {\bibfnamefont {K.}~\bibnamefont
  {Olejnik}},\ }\bibfield  {title} {\enquote {\bibinfo {title} {Spin {Hall}
  effect devices},}\ }\href {\doibase 10.1038/nmat3279} {\bibfield  {journal}
  {\bibinfo  {journal} {Nat. Mater.}\ }\textbf {\bibinfo {volume} {11}},\
  \bibinfo {pages} {382} (\bibinfo {year} {2012})}\BibitemShut {NoStop}%
\bibitem [{\citenamefont {Manchon}\ \emph {et~al.}(2015)\citenamefont
  {Manchon}, \citenamefont {Koo}, \citenamefont {Nitta}, \citenamefont
  {Frolov},\ and\ \citenamefont {Duine}}]{manchon.koo.15}%
  \BibitemOpen
  \bibfield  {author} {\bibinfo {author} {\bibfnamefont {A.}~\bibnamefont
  {Manchon}}, \bibinfo {author} {\bibfnamefont {H.~C.}\ \bibnamefont {Koo}},
  \bibinfo {author} {\bibfnamefont {J.}~\bibnamefont {Nitta}}, \bibinfo
  {author} {\bibfnamefont {S.~M.}\ \bibnamefont {Frolov}}, \ and\ \bibinfo
  {author} {\bibfnamefont {R.A.}\ \bibnamefont {Duine}},\ }\bibfield  {title}
  {\enquote {\bibinfo {title} {New perspectives for {Rashba} spin-orbit
  coupling},}\ }\href {\doibase 10.1038/nmat4360} {\bibfield  {journal}
  {\bibinfo  {journal} {Nat. Mater.}\ }\textbf {\bibinfo {volume} {14}},\
  \bibinfo {pages} {871} (\bibinfo {year} {2015})}\BibitemShut {NoStop}%
\bibitem [{\citenamefont {Joshi}(2016)}]{joshi.16}%
  \BibitemOpen
  \bibfield  {author} {\bibinfo {author} {\bibfnamefont {V.~K.}\ \bibnamefont
  {Joshi}},\ }\bibfield  {title} {\enquote {\bibinfo {title} {Spintronics: A
  contemporary review of emerging electronics devices},}\ }\href {\doibase
  10.1016/j.jestch.2016.05.002} {\bibfield  {journal} {\bibinfo  {journal}
  {Engineering Science and Technology, an International Journal (ESTIJ)}\
  }\textbf {\bibinfo {volume} {19}},\ \bibinfo {pages} {1503} (\bibinfo {year}
  {2016})}\BibitemShut {NoStop}%
\bibitem [{\citenamefont {Soumyanarayanan}\ \emph {et~al.}(2016)\citenamefont
  {Soumyanarayanan}, \citenamefont {Reyren}, \citenamefont {Fert},\ and\
  \citenamefont {Panagopoulos}}]{soumyanarayanan.reyren.16}%
  \BibitemOpen
  \bibfield  {author} {\bibinfo {author} {\bibfnamefont {A.}~\bibnamefont
  {Soumyanarayanan}}, \bibinfo {author} {\bibfnamefont {N.}~\bibnamefont
  {Reyren}}, \bibinfo {author} {\bibfnamefont {A.}~\bibnamefont {Fert}}, \ and\
  \bibinfo {author} {\bibfnamefont {Ch.}\ \bibnamefont {Panagopoulos}},\
  }\bibfield  {title} {\enquote {\bibinfo {title} {Emergent phenomena induced
  by spin-orbit coupling at surfaces and interfaces},}\ }\href {\doibase
  10.1038/nature19820} {\bibfield  {journal} {\bibinfo  {journal} {Nature}\
  }\textbf {\bibinfo {volume} {539}},\ \bibinfo {pages} {509} (\bibinfo {year}
  {2016})}\BibitemShut {NoStop}%
\bibitem [{\citenamefont {Chappert}\ \emph {et~al.}(2007)\citenamefont
  {Chappert}, \citenamefont {Fert},\ and\ \citenamefont
  {Van~Dau}}]{chappert.fert.07}%
  \BibitemOpen
  \bibfield  {author} {\bibinfo {author} {\bibfnamefont {C.}~\bibnamefont
  {Chappert}}, \bibinfo {author} {\bibfnamefont {A.}~\bibnamefont {Fert}}, \
  and\ \bibinfo {author} {\bibfnamefont {F.~N.}\ \bibnamefont {Van~Dau}},\
  }\bibfield  {title} {\enquote {\bibinfo {title} {The emergence of spin
  electronics in data storage},}\ }\href {\doibase 10.1038/nmat2024} {\bibfield
   {journal} {\bibinfo  {journal} {Nat. Mater.}\ }\textbf {\bibinfo {volume}
  {6}},\ \bibinfo {pages} {813} (\bibinfo {year} {2007})}\BibitemShut {NoStop}%
\bibitem [{\citenamefont {Knill}(2005)}]{knill.05}%
  \BibitemOpen
  \bibfield  {author} {\bibinfo {author} {\bibfnamefont {E.}~\bibnamefont
  {Knill}},\ }\bibfield  {title} {\enquote {\bibinfo {title} {Quantum computing
  with realistically noisy devices},}\ }\href {\doibase 10.1038/nature03350}
  {\bibfield  {journal} {\bibinfo  {journal} {Nature}\ }\textbf {\bibinfo
  {volume} {434}},\ \bibinfo {pages} {39} (\bibinfo {year} {2005})}\BibitemShut
  {NoStop}%
\bibitem [{\citenamefont {Friedman}\ \emph {et~al.}(2015)\citenamefont
  {Friedman}, \citenamefont {Fadel}, \citenamefont {Wessels}, \citenamefont
  {Querlioz},\ and\ \citenamefont {Sahakian}}]{joseph.eric.15}%
  \BibitemOpen
  \bibfield  {author} {\bibinfo {author} {\bibfnamefont {J.~S.}\ \bibnamefont
  {Friedman}}, \bibinfo {author} {\bibfnamefont {E.~R.}\ \bibnamefont {Fadel}},
  \bibinfo {author} {\bibfnamefont {B.~W.}\ \bibnamefont {Wessels}}, \bibinfo
  {author} {\bibfnamefont {D.}~\bibnamefont {Querlioz}}, \ and\ \bibinfo
  {author} {\bibfnamefont {A.~V.}\ \bibnamefont {Sahakian}},\ }\bibfield
  {title} {\enquote {\bibinfo {title} {Bilayer avalanche spin-diode logic},}\
  }\href {\doibase 10.1063/1.4935262} {\bibfield  {journal} {\bibinfo
  {journal} {AIP Advances}\ }\textbf {\bibinfo {volume} {5}},\ \bibinfo {pages}
  {117102} (\bibinfo {year} {2015})}\BibitemShut {NoStop}%
\bibitem [{\citenamefont {Wang}\ \emph {et~al.}(2016)\citenamefont {Wang},
  \citenamefont {Kumar}, \citenamefont {Wu},\ and\ \citenamefont
  {Weiss}}]{wang.kumar.16}%
  \BibitemOpen
  \bibfield  {author} {\bibinfo {author} {\bibfnamefont {Y.}~\bibnamefont
  {Wang}}, \bibinfo {author} {\bibfnamefont {A.}~\bibnamefont {Kumar}},
  \bibinfo {author} {\bibfnamefont {T.-Y.}\ \bibnamefont {Wu}}, \ and\ \bibinfo
  {author} {\bibfnamefont {D.~S.}\ \bibnamefont {Weiss}},\ }\bibfield  {title}
  {\enquote {\bibinfo {title} {Single-qubit gates based on targeted phase
  shifts in a {3D} neutral atom array},}\ }\href {\doibase
  10.1126/science.aaf2581} {\bibfield  {journal} {\bibinfo  {journal}
  {Science}\ }\textbf {\bibinfo {volume} {352}},\ \bibinfo {pages} {1562}
  (\bibinfo {year} {2016})}\BibitemShut {NoStop}%
\bibitem [{\citenamefont {Linder}\ and\ \citenamefont
  {Robinson}(2015)}]{linder.robinson.15}%
  \BibitemOpen
  \bibfield  {author} {\bibinfo {author} {\bibfnamefont {J.}~\bibnamefont
  {Linder}}\ and\ \bibinfo {author} {\bibfnamefont {J.~W.~A.}\ \bibnamefont
  {Robinson}},\ }\bibfield  {title} {\enquote {\bibinfo {title}
  {Superconducting spintronics},}\ }\href {\doibase 10.1038/nphys3242}
  {\bibfield  {journal} {\bibinfo  {journal} {Nat. Phys.}\ }\textbf {\bibinfo
  {volume} {11}},\ \bibinfo {pages} {307} (\bibinfo {year} {2015})}\BibitemShut
  {NoStop}%
\bibitem [{\citenamefont {Gingrich}\ \emph {et~al.}(2016)\citenamefont
  {Gingrich}, \citenamefont {Niedzielski}, \citenamefont {Glick}, \citenamefont
  {Wang}, \citenamefont {Miller}, \citenamefont {Loloee}, \citenamefont
  {Pratt~Jr},\ and\ \citenamefont {Birge}}]{gingrich.niedzielski.16}%
  \BibitemOpen
  \bibfield  {author} {\bibinfo {author} {\bibfnamefont {E.~C.}\ \bibnamefont
  {Gingrich}}, \bibinfo {author} {\bibfnamefont {B.~M.}\ \bibnamefont
  {Niedzielski}}, \bibinfo {author} {\bibfnamefont {J.~A.}\ \bibnamefont
  {Glick}}, \bibinfo {author} {\bibfnamefont {Y.}~\bibnamefont {Wang}},
  \bibinfo {author} {\bibfnamefont {D.~L.}\ \bibnamefont {Miller}}, \bibinfo
  {author} {\bibfnamefont {R.}~\bibnamefont {Loloee}}, \bibinfo {author}
  {\bibfnamefont {W.~P.}\ \bibnamefont {Pratt~Jr}}, \ and\ \bibinfo {author}
  {\bibfnamefont {N.~O.}\ \bibnamefont {Birge}},\ }\bibfield  {title} {\enquote
  {\bibinfo {title} {Controllable 0-$\pi$ {Josephson} junctions containing a
  ferromagnetic spin valve},}\ }\href {\doibase 10.1038/nphys3681} {\bibfield
  {journal} {\bibinfo  {journal} {Nat. Phys.}\ }\textbf {\bibinfo {volume}
  {12}},\ \bibinfo {pages} {564} (\bibinfo {year} {2016})}\BibitemShut
  {NoStop}%
\bibitem [{\citenamefont {Costa}\ \emph {et~al.}(2017)\citenamefont {Costa},
  \citenamefont {H\"{o}gl},\ and\ \citenamefont {Fabian}}]{costa.hogl.17}%
  \BibitemOpen
  \bibfield  {author} {\bibinfo {author} {\bibfnamefont {A.}~\bibnamefont
  {Costa}}, \bibinfo {author} {\bibfnamefont {P.}~\bibnamefont {H\"{o}gl}}, \
  and\ \bibinfo {author} {\bibfnamefont {J.}~\bibnamefont {Fabian}},\
  }\bibfield  {title} {\enquote {\bibinfo {title} {Magnetoanisotropic
  {Josephson} effect due to interfacial spin-orbit fields in
  superconductor/ferromagnet/superconductor junctions},}\ }\href {\doibase
  10.1103/PhysRevB.95.024514} {\bibfield  {journal} {\bibinfo  {journal} {Phys.
  Rev. B}\ }\textbf {\bibinfo {volume} {95}},\ \bibinfo {pages} {024514}
  (\bibinfo {year} {2017})}\BibitemShut {NoStop}%
\bibitem [{\citenamefont {Niu}(2012)}]{niu.12}%
  \BibitemOpen
  \bibfield  {author} {\bibinfo {author} {\bibfnamefont {Z.-P.}\ \bibnamefont
  {Niu}},\ }\bibfield  {title} {\enquote {\bibinfo {title} {A spin triplet
  supercurrent in half metal ferromagnet/superconductor junctions with the
  interfacial {Rashba} spin-orbit coupling},}\ }\href {\doibase
  10.1063/1.4743001} {\bibfield  {journal} {\bibinfo  {journal} {Appl. Phys.
  Lett.}\ }\textbf {\bibinfo {volume} {101}},\ \bibinfo {pages} {062601}
  (\bibinfo {year} {2012})}\BibitemShut {NoStop}%
\bibitem [{\citenamefont {Alicea}(2012)}]{alicea.12}%
  \BibitemOpen
  \bibfield  {author} {\bibinfo {author} {\bibfnamefont {J.}~\bibnamefont
  {Alicea}},\ }\bibfield  {title} {\enquote {\bibinfo {title} {New directions
  in the pursuit of {Majorana} fermions in solid state systems},}\ }\href
  {\doibase 10.1088/0034-4885/75/7/076501} {\bibfield  {journal} {\bibinfo
  {journal} {Rep. Prog. Phys.}\ }\textbf {\bibinfo {volume} {75}},\ \bibinfo
  {pages} {076501} (\bibinfo {year} {2012})}\BibitemShut {NoStop}%
\bibitem [{\citenamefont {Beenakker}(2013)}]{beenakker.13}%
  \BibitemOpen
  \bibfield  {author} {\bibinfo {author} {\bibfnamefont {C.~W.~J.}\
  \bibnamefont {Beenakker}},\ }\bibfield  {title} {\enquote {\bibinfo {title}
  {Search for {Majorana} fermions in superconductors},}\ }\href {\doibase
  10.1146/annurev-conmatphys-030212-184337} {\bibfield  {journal} {\bibinfo
  {journal} {Annu. Rev. Condens. Matter Phys.}\ }\textbf {\bibinfo {volume}
  {4}},\ \bibinfo {pages} {113} (\bibinfo {year} {2013})}\BibitemShut {NoStop}%
\bibitem [{\citenamefont {Elliott}\ and\ \citenamefont
  {Franz}(2015)}]{eliot.franz.15}%
  \BibitemOpen
  \bibfield  {author} {\bibinfo {author} {\bibfnamefont {S.~R.}\ \bibnamefont
  {Elliott}}\ and\ \bibinfo {author} {\bibfnamefont {M.}~\bibnamefont
  {Franz}},\ }\bibfield  {title} {\enquote {\bibinfo {title} {{\it Colloquium}:
  Majorana fermions in nuclear, particle, and solid-state physics},}\ }\href
  {\doibase 10.1103/RevModPhys.87.137} {\bibfield  {journal} {\bibinfo
  {journal} {Rev. Mod. Phys.}\ }\textbf {\bibinfo {volume} {87}},\ \bibinfo
  {pages} {137} (\bibinfo {year} {2015})}\BibitemShut {NoStop}%
\bibitem [{\citenamefont {Kitaev}(2001)}]{kitaev.01}%
  \BibitemOpen
  \bibfield  {author} {\bibinfo {author} {\bibfnamefont {A.~Y.}\ \bibnamefont
  {Kitaev}},\ }\bibfield  {title} {\enquote {\bibinfo {title} {Unpaired
  {Majorana} fermions in quantum wires},}\ }\href {\doibase
  10.1070/1063-7869/44/10S/S29} {\bibfield  {journal} {\bibinfo  {journal}
  {Phys.-Usp.}\ }\textbf {\bibinfo {volume} {44}},\ \bibinfo {pages} {131}
  (\bibinfo {year} {2001})}\BibitemShut {NoStop}%
\bibitem [{\citenamefont {Mourik}\ \emph {et~al.}(2012)\citenamefont {Mourik},
  \citenamefont {Zuo}, \citenamefont {Frolov}, \citenamefont {Plissard},
  \citenamefont {Bakkers},\ and\ \citenamefont {Kouwenhoven}}]{mourik.zuo.12}%
  \BibitemOpen
  \bibfield  {author} {\bibinfo {author} {\bibfnamefont {V.}~\bibnamefont
  {Mourik}}, \bibinfo {author} {\bibfnamefont {K.}~\bibnamefont {Zuo}},
  \bibinfo {author} {\bibfnamefont {S.~M.}\ \bibnamefont {Frolov}}, \bibinfo
  {author} {\bibfnamefont {S.~R.}\ \bibnamefont {Plissard}}, \bibinfo {author}
  {\bibfnamefont {E.~P. A.~M.}\ \bibnamefont {Bakkers}}, \ and\ \bibinfo
  {author} {\bibfnamefont {L.~P.}\ \bibnamefont {Kouwenhoven}},\ }\bibfield
  {title} {\enquote {\bibinfo {title} {Signatures of {Majorana} fermions in
  hybrid superconductor-semiconductor nanowire devices},}\ }\href {\doibase
  10.1126/science.1222360} {\bibfield  {journal} {\bibinfo  {journal}
  {Science}\ }\textbf {\bibinfo {volume} {336}},\ \bibinfo {pages} {1003}
  (\bibinfo {year} {2012})}\BibitemShut {NoStop}%
\bibitem [{\citenamefont {Das}\ \emph {et~al.}(2012)\citenamefont {Das},
  \citenamefont {Ronen}, \citenamefont {Most}, \citenamefont {Oreg},
  \citenamefont {Heiblum},\ and\ \citenamefont {Shtrikman}}]{das.ronen.12}%
  \BibitemOpen
  \bibfield  {author} {\bibinfo {author} {\bibfnamefont {A.}~\bibnamefont
  {Das}}, \bibinfo {author} {\bibfnamefont {Y.}~\bibnamefont {Ronen}}, \bibinfo
  {author} {\bibfnamefont {Y.}~\bibnamefont {Most}}, \bibinfo {author}
  {\bibfnamefont {Y.}~\bibnamefont {Oreg}}, \bibinfo {author} {\bibfnamefont
  {M.}~\bibnamefont {Heiblum}}, \ and\ \bibinfo {author} {\bibfnamefont
  {H.}~\bibnamefont {Shtrikman}},\ }\bibfield  {title} {\enquote {\bibinfo
  {title} {Zero-bias peaks and splitting in an {Al-InAs} nanowire topological
  superconductor as a signature of {Majorana} fermions},}\ }\href {\doibase
  10.1038/nphys2479} {\bibfield  {journal} {\bibinfo  {journal} {Nat. Phys.}\
  }\textbf {\bibinfo {volume} {8}},\ \bibinfo {pages} {887} (\bibinfo {year}
  {2012})}\BibitemShut {NoStop}%
\bibitem [{\citenamefont {Deng}\ \emph {et~al.}(2012)\citenamefont {Deng},
  \citenamefont {Yu}, \citenamefont {Huang}, \citenamefont {Larsson},
  \citenamefont {Caroff},\ and\ \citenamefont {Xu}}]{deng.yu.12}%
  \BibitemOpen
  \bibfield  {author} {\bibinfo {author} {\bibfnamefont {M.~T.}\ \bibnamefont
  {Deng}}, \bibinfo {author} {\bibfnamefont {C.~L.}\ \bibnamefont {Yu}},
  \bibinfo {author} {\bibfnamefont {G.~Y.}\ \bibnamefont {Huang}}, \bibinfo
  {author} {\bibfnamefont {M.}~\bibnamefont {Larsson}}, \bibinfo {author}
  {\bibfnamefont {P.}~\bibnamefont {Caroff}}, \ and\ \bibinfo {author}
  {\bibfnamefont {H.~Q.}\ \bibnamefont {Xu}},\ }\bibfield  {title} {\enquote
  {\bibinfo {title} {Anomalous zero-bias conductance peak in a {Nb-InSb}
  nanowire-{Nb} hybrid device},}\ }\href {\doibase 10.1021/nl303758w}
  {\bibfield  {journal} {\bibinfo  {journal} {Nano Lett.}\ }\textbf {\bibinfo
  {volume} {12}},\ \bibinfo {pages} {6414} (\bibinfo {year}
  {2012})}\BibitemShut {NoStop}%
\bibitem [{\citenamefont {Churchill}\ \emph {et~al.}(2013)\citenamefont
  {Churchill}, \citenamefont {Fatemi}, \citenamefont {Grove-Rasmussen},
  \citenamefont {Deng}, \citenamefont {Caroff}, \citenamefont {Xu},\ and\
  \citenamefont {Marcus}}]{churchill.fatemi.13}%
  \BibitemOpen
  \bibfield  {author} {\bibinfo {author} {\bibfnamefont {H.~O.~H.}\
  \bibnamefont {Churchill}}, \bibinfo {author} {\bibfnamefont {V.}~\bibnamefont
  {Fatemi}}, \bibinfo {author} {\bibfnamefont {K.}~\bibnamefont
  {Grove-Rasmussen}}, \bibinfo {author} {\bibfnamefont {M.~T.}\ \bibnamefont
  {Deng}}, \bibinfo {author} {\bibfnamefont {P.}~\bibnamefont {Caroff}},
  \bibinfo {author} {\bibfnamefont {H.~Q.}\ \bibnamefont {Xu}}, \ and\ \bibinfo
  {author} {\bibfnamefont {C.~M.}\ \bibnamefont {Marcus}},\ }\bibfield  {title}
  {\enquote {\bibinfo {title} {Superconductor-nanowire devices from tunneling
  to the multichannel regime: Zero-bias oscillations and magnetoconductance
  crossover},}\ }\href {\doibase 10.1103/PhysRevB.87.241401} {\bibfield
  {journal} {\bibinfo  {journal} {Phys. Rev. B}\ }\textbf {\bibinfo {volume}
  {87}},\ \bibinfo {pages} {241401} (\bibinfo {year} {2013})}\BibitemShut
  {NoStop}%
\bibitem [{\citenamefont {Feldman}\ \emph {et~al.}(2017)\citenamefont
  {Feldman}, \citenamefont {Randeria}, \citenamefont {Li}, \citenamefont
  {Jeon}, \citenamefont {Xie}, \citenamefont {Wang}, \citenamefont {Drozdov},
  \citenamefont {Andrei~B.},\ and\ \citenamefont
  {Yazdani}}]{feldman.randeria.17}%
  \BibitemOpen
  \bibfield  {author} {\bibinfo {author} {\bibfnamefont {B.~E.}\ \bibnamefont
  {Feldman}}, \bibinfo {author} {\bibfnamefont {M.~T.}\ \bibnamefont
  {Randeria}}, \bibinfo {author} {\bibfnamefont {J.}~\bibnamefont {Li}},
  \bibinfo {author} {\bibfnamefont {S.}~\bibnamefont {Jeon}}, \bibinfo {author}
  {\bibfnamefont {Y.}~\bibnamefont {Xie}}, \bibinfo {author} {\bibfnamefont
  {Z.}~\bibnamefont {Wang}}, \bibinfo {author} {\bibfnamefont {I.~K.}\
  \bibnamefont {Drozdov}}, \bibinfo {author} {\bibfnamefont {B.}~\bibnamefont
  {Andrei~B.}}, \ and\ \bibinfo {author} {\bibfnamefont {A.}~\bibnamefont
  {Yazdani}},\ }\bibfield  {title} {\enquote {\bibinfo {title} {High-resolution
  studies of the {Majorana} atomic chain platform},}\ }\href {\doibase
  10.1038/nphys3947} {\bibfield  {journal} {\bibinfo  {journal} {Nat. Phys.}\
  }\textbf {\bibinfo {volume} {13}},\ \bibinfo {pages} {286} (\bibinfo {year}
  {2017})}\BibitemShut {NoStop}%
\bibitem [{\citenamefont {Deng}\ \emph {et~al.}(2016)\citenamefont {Deng},
  \citenamefont {Vaitiekenas}, \citenamefont {Hansen}, \citenamefont {Danon},
  \citenamefont {Leijnse}, \citenamefont {Flensberg}, \citenamefont
  {Nyg\aa{}rd}, \citenamefont {Krogstrup},\ and\ \citenamefont
  {Marcus}}]{deng.vaitiekenas.16}%
  \BibitemOpen
  \bibfield  {author} {\bibinfo {author} {\bibfnamefont {M.~T.}\ \bibnamefont
  {Deng}}, \bibinfo {author} {\bibfnamefont {S.}~\bibnamefont {Vaitiekenas}},
  \bibinfo {author} {\bibfnamefont {E.~B.}\ \bibnamefont {Hansen}}, \bibinfo
  {author} {\bibfnamefont {J.}~\bibnamefont {Danon}}, \bibinfo {author}
  {\bibfnamefont {M.}~\bibnamefont {Leijnse}}, \bibinfo {author} {\bibfnamefont
  {K.}~\bibnamefont {Flensberg}}, \bibinfo {author} {\bibfnamefont
  {J.}~\bibnamefont {Nyg\aa{}rd}}, \bibinfo {author} {\bibfnamefont
  {P.}~\bibnamefont {Krogstrup}}, \ and\ \bibinfo {author} {\bibfnamefont
  {C.~M.}\ \bibnamefont {Marcus}},\ }\bibfield  {title} {\enquote {\bibinfo
  {title} {Majorana bound state in a coupled quantum-dot hybrid-nanowire
  system},}\ }\href {\doibase 10.1126/science.aaf3961} {\bibfield  {journal}
  {\bibinfo  {journal} {Science}\ }\textbf {\bibinfo {volume} {354}},\ \bibinfo
  {pages} {1557} (\bibinfo {year} {2016})}\BibitemShut {NoStop}%
\bibitem [{\citenamefont {Ptok}\ \emph {et~al.}(2017)\citenamefont {Ptok},
  \citenamefont {Kobia\l{}ka},\ and\ \citenamefont
  {Doma\'{n}ski}}]{ptok.kobialka.17}%
  \BibitemOpen
  \bibfield  {author} {\bibinfo {author} {\bibfnamefont {A.}~\bibnamefont
  {Ptok}}, \bibinfo {author} {\bibfnamefont {A.}~\bibnamefont {Kobia\l{}ka}}, \
  and\ \bibinfo {author} {\bibfnamefont {T.}~\bibnamefont {Doma\'{n}ski}},\
  }\bibfield  {title} {\enquote {\bibinfo {title} {Controlling the bound states
  in a quantum-dot hybrid nanowire},}\ }\href {\doibase
  10.1103/PhysRevB.96.195430} {\bibfield  {journal} {\bibinfo  {journal} {Phys.
  Rev. B}\ }\textbf {\bibinfo {volume} {96}},\ \bibinfo {pages} {195430}
  (\bibinfo {year} {2017})}\BibitemShut {NoStop}%
\bibitem [{\citenamefont {Fu}\ and\ \citenamefont {Kane}(2008)}]{fu.kane.08}%
  \BibitemOpen
  \bibfield  {author} {\bibinfo {author} {\bibfnamefont {L.}~\bibnamefont
  {Fu}}\ and\ \bibinfo {author} {\bibfnamefont {C.~L.}\ \bibnamefont {Kane}},\
  }\bibfield  {title} {\enquote {\bibinfo {title} {Superconducting proximity
  effect and {Majorana} fermions at the surface of a topological insulator},}\
  }\href {\doibase 10.1103/PhysRevLett.100.096407} {\bibfield  {journal}
  {\bibinfo  {journal} {Phys. Rev. Lett.}\ }\textbf {\bibinfo {volume} {100}},\
  \bibinfo {pages} {096407} (\bibinfo {year} {2008})}\BibitemShut {NoStop}%
\bibitem [{\citenamefont {Tewari}\ \emph {et~al.}(2007)\citenamefont {Tewari},
  \citenamefont {Das~Sarma}, \citenamefont {Nayak}, \citenamefont {Zhang},\
  and\ \citenamefont {Zoller}}]{tewari.dassarma.07}%
  \BibitemOpen
  \bibfield  {author} {\bibinfo {author} {\bibfnamefont {S.}~\bibnamefont
  {Tewari}}, \bibinfo {author} {\bibfnamefont {S.}~\bibnamefont {Das~Sarma}},
  \bibinfo {author} {\bibfnamefont {Chetan}\ \bibnamefont {Nayak}}, \bibinfo
  {author} {\bibfnamefont {Ch.}\ \bibnamefont {Zhang}}, \ and\ \bibinfo
  {author} {\bibfnamefont {P.}~\bibnamefont {Zoller}},\ }\bibfield  {title}
  {\enquote {\bibinfo {title} {Quantum computation using vortices and
  {Majorana} zero modes of a ${p}_{x}+i{p}_{y}$ superfluid of fermionic cold
  atoms},}\ }\href {\doibase 10.1103/PhysRevLett.98.010506} {\bibfield
  {journal} {\bibinfo  {journal} {Phys. Rev. Lett.}\ }\textbf {\bibinfo
  {volume} {98}},\ \bibinfo {pages} {010506} (\bibinfo {year}
  {2007})}\BibitemShut {NoStop}%
\bibitem [{\citenamefont {Hsu}\ \emph {et~al.}(2008)\citenamefont {Hsu},
  \citenamefont {Luo}, \citenamefont {Yeh}, \citenamefont {Chen}, \citenamefont
  {Huang}, \citenamefont {Wu}, \citenamefont {Lee}, \citenamefont {Huang},
  \citenamefont {Chu}, \citenamefont {Yan},\ and\ \citenamefont
  {Wu}}]{hsu.luo.08}%
  \BibitemOpen
  \bibfield  {author} {\bibinfo {author} {\bibfnamefont {F.-Ch.}\ \bibnamefont
  {Hsu}}, \bibinfo {author} {\bibfnamefont {J.-Y.}\ \bibnamefont {Luo}},
  \bibinfo {author} {\bibfnamefont {K.-W.}\ \bibnamefont {Yeh}}, \bibinfo
  {author} {\bibfnamefont {T.-K.}\ \bibnamefont {Chen}}, \bibinfo {author}
  {\bibfnamefont {T.-W.}\ \bibnamefont {Huang}}, \bibinfo {author}
  {\bibfnamefont {P.~M.}\ \bibnamefont {Wu}}, \bibinfo {author} {\bibfnamefont
  {Y.-Ch..}\ \bibnamefont {Lee}}, \bibinfo {author} {\bibfnamefont {Y.-L.}\
  \bibnamefont {Huang}}, \bibinfo {author} {\bibfnamefont {Y.-Y.}\ \bibnamefont
  {Chu}}, \bibinfo {author} {\bibfnamefont {D.-Ch.}\ \bibnamefont {Yan}}, \
  and\ \bibinfo {author} {\bibfnamefont {M.-K.}\ \bibnamefont {Wu}},\
  }\bibfield  {title} {\enquote {\bibinfo {title} {Superconductivity in the
  {PbO}-type structure {$\alpha$-FeSe}},}\ }\href {\doibase
  10.1073/pnas.0807325105} {\bibfield  {journal} {\bibinfo  {journal} {PNAS}\
  }\textbf {\bibinfo {volume} {105}},\ \bibinfo {pages} {14262} (\bibinfo
  {year} {2008})}\BibitemShut {NoStop}%
\bibitem [{\citenamefont {Qing-Yan}\ \emph {et~al.}(2012)\citenamefont
  {Qing-Yan}, \citenamefont {Zhi}, \citenamefont {Wen-Hao}, \citenamefont
  {Zuo-Cheng}, \citenamefont {Jin-Song}, \citenamefont {Wei}, \citenamefont
  {Hao}, \citenamefont {Yun-Bo}, \citenamefont {Peng}, \citenamefont {Kai},
  \citenamefont {Jing}, \citenamefont {Can-Li}, \citenamefont {Ke},
  \citenamefont {Jin-Feng}, \citenamefont {Shuai-Hua}, \citenamefont {Ya-Yu},
  \citenamefont {Li-Li}, \citenamefont {Xi}, \citenamefont {Xu-Cun},\ and\
  \citenamefont {Qi-Kun}}]{qingyan.zhi.12}%
  \BibitemOpen
  \bibfield  {author} {\bibinfo {author} {\bibfnamefont {W.}~\bibnamefont
  {Qing-Yan}}, \bibinfo {author} {\bibfnamefont {L.}~\bibnamefont {Zhi}},
  \bibinfo {author} {\bibfnamefont {Z.}~\bibnamefont {Wen-Hao}}, \bibinfo
  {author} {\bibfnamefont {Z.}~\bibnamefont {Zuo-Cheng}}, \bibinfo {author}
  {\bibfnamefont {Z.}~\bibnamefont {Jin-Song}}, \bibinfo {author}
  {\bibfnamefont {L.}~\bibnamefont {Wei}}, \bibinfo {author} {\bibfnamefont
  {D.}~\bibnamefont {Hao}}, \bibinfo {author} {\bibfnamefont {O.}~\bibnamefont
  {Yun-Bo}}, \bibinfo {author} {\bibfnamefont {D.}~\bibnamefont {Peng}},
  \bibinfo {author} {\bibfnamefont {Ch.}\ \bibnamefont {Kai}}, \bibinfo
  {author} {\bibfnamefont {W.}~\bibnamefont {Jing}}, \bibinfo {author}
  {\bibfnamefont {S.}~\bibnamefont {Can-Li}}, \bibinfo {author} {\bibfnamefont
  {H.}~\bibnamefont {Ke}}, \bibinfo {author} {\bibfnamefont {J.}~\bibnamefont
  {Jin-Feng}}, \bibinfo {author} {\bibfnamefont {J.}~\bibnamefont {Shuai-Hua}},
  \bibinfo {author} {\bibfnamefont {W.}~\bibnamefont {Ya-Yu}}, \bibinfo
  {author} {\bibfnamefont {W.}~\bibnamefont {Li-Li}}, \bibinfo {author}
  {\bibfnamefont {Ch.}\ \bibnamefont {Xi}}, \bibinfo {author} {\bibfnamefont
  {M.}~\bibnamefont {Xu-Cun}}, \ and\ \bibinfo {author} {\bibfnamefont
  {X.}~\bibnamefont {Qi-Kun}},\ }\bibfield  {title} {\enquote {\bibinfo {title}
  {Interface-induced high-temperature superconductivity in single unit-cell
  {FeSe} films on {SrTiO$_{3}$}},}\ }\href {\doibase
  10.1088/0256-307X/29/3/037402} {\bibfield  {journal} {\bibinfo  {journal}
  {Chinese Phys. Lett.}\ }\textbf {\bibinfo {volume} {29}},\ \bibinfo {pages}
  {037402} (\bibinfo {year} {2012})}\BibitemShut {NoStop}%
\bibitem [{\citenamefont {Liu}\ \emph {et~al.}(2012)\citenamefont {Liu},
  \citenamefont {Zhang}, \citenamefont {Mou}, \citenamefont {He}, \citenamefont
  {Ou}, \citenamefont {Wang}, \citenamefont {Li}, \citenamefont {Wang},
  \citenamefont {Zhao}, \citenamefont {He}, \citenamefont {Peng}, \citenamefont
  {Liu}, \citenamefont {Chen}, \citenamefont {Yu}, \citenamefont {Liu},
  \citenamefont {Dong}, \citenamefont {Zhang}, \citenamefont {Chen},
  \citenamefont {Xu}, \citenamefont {Hu}, \citenamefont {Chen}, \citenamefont
  {Ma}, \citenamefont {Xue},\ and\ \citenamefont {Zhou}}]{liu.zhang.12}%
  \BibitemOpen
  \bibfield  {author} {\bibinfo {author} {\bibfnamefont {D.}~\bibnamefont
  {Liu}}, \bibinfo {author} {\bibfnamefont {W.}~\bibnamefont {Zhang}}, \bibinfo
  {author} {\bibfnamefont {D.}~\bibnamefont {Mou}}, \bibinfo {author}
  {\bibfnamefont {J.}~\bibnamefont {He}}, \bibinfo {author} {\bibfnamefont
  {Y.-B.}\ \bibnamefont {Ou}}, \bibinfo {author} {\bibfnamefont {Q.-Y.}\
  \bibnamefont {Wang}}, \bibinfo {author} {\bibfnamefont {Z.}~\bibnamefont
  {Li}}, \bibinfo {author} {\bibfnamefont {L.}~\bibnamefont {Wang}}, \bibinfo
  {author} {\bibfnamefont {L.}~\bibnamefont {Zhao}}, \bibinfo {author}
  {\bibfnamefont {S.}~\bibnamefont {He}}, \bibinfo {author} {\bibfnamefont
  {Y.}~\bibnamefont {Peng}}, \bibinfo {author} {\bibfnamefont {X.}~\bibnamefont
  {Liu}}, \bibinfo {author} {\bibfnamefont {Ch.}\ \bibnamefont {Chen}},
  \bibinfo {author} {\bibfnamefont {L.}~\bibnamefont {Yu}}, \bibinfo {author}
  {\bibfnamefont {G.}~\bibnamefont {Liu}}, \bibinfo {author} {\bibfnamefont
  {X.}~\bibnamefont {Dong}}, \bibinfo {author} {\bibfnamefont {J.}~\bibnamefont
  {Zhang}}, \bibinfo {author} {\bibfnamefont {Ch.}\ \bibnamefont {Chen}},
  \bibinfo {author} {\bibfnamefont {Z.}~\bibnamefont {Xu}}, \bibinfo {author}
  {\bibfnamefont {J.}~\bibnamefont {Hu}}, \bibinfo {author} {\bibfnamefont
  {X.}~\bibnamefont {Chen}}, \bibinfo {author} {\bibfnamefont {X.}~\bibnamefont
  {Ma}}, \bibinfo {author} {\bibfnamefont {Q.}~\bibnamefont {Xue}}, \ and\
  \bibinfo {author} {\bibfnamefont {X.~J.}\ \bibnamefont {Zhou}},\ }\bibfield
  {title} {\enquote {\bibinfo {title} {Electronic origin of high-temperature
  superconductivity in single-layer {FeSe} superconductor},}\ }\href {\doibase
  10.1038/ncomms1946} {\bibfield  {journal} {\bibinfo  {journal} {Nat.
  Commun.}\ }\textbf {\bibinfo {volume} {3}},\ \bibinfo {pages} {931} (\bibinfo
  {year} {2012})}\BibitemShut {NoStop}%
\bibitem [{\citenamefont {He}\ \emph {et~al.}(2013)\citenamefont {He},
  \citenamefont {He}, \citenamefont {Zhang}, \citenamefont {Zhao},
  \citenamefont {Liu}, \citenamefont {Liu}, \citenamefont {Mou}, \citenamefont
  {Ou}, \citenamefont {Wang}, \citenamefont {Li}, \citenamefont {Wang},
  \citenamefont {Peng}, \citenamefont {Liu}, \citenamefont {Chen},
  \citenamefont {Yu}, \citenamefont {Liu}, \citenamefont {Dong}, \citenamefont
  {Zhang}, \citenamefont {Chen}, \citenamefont {Xu}, \citenamefont {Chen},
  \citenamefont {Ma}, \citenamefont {Xue},\ and\ \citenamefont
  {Zhou}}]{he.he.13}%
  \BibitemOpen
  \bibfield  {author} {\bibinfo {author} {\bibfnamefont {S.}~\bibnamefont
  {He}}, \bibinfo {author} {\bibfnamefont {J.}~\bibnamefont {He}}, \bibinfo
  {author} {\bibfnamefont {W.}~\bibnamefont {Zhang}}, \bibinfo {author}
  {\bibfnamefont {L.}~\bibnamefont {Zhao}}, \bibinfo {author} {\bibfnamefont
  {D.}~\bibnamefont {Liu}}, \bibinfo {author} {\bibfnamefont {X.}~\bibnamefont
  {Liu}}, \bibinfo {author} {\bibfnamefont {D.}~\bibnamefont {Mou}}, \bibinfo
  {author} {\bibfnamefont {Y.-B.}\ \bibnamefont {Ou}}, \bibinfo {author}
  {\bibfnamefont {Q.-Y.}\ \bibnamefont {Wang}}, \bibinfo {author}
  {\bibfnamefont {Z.}~\bibnamefont {Li}}, \bibinfo {author} {\bibfnamefont
  {L.}~\bibnamefont {Wang}}, \bibinfo {author} {\bibfnamefont {Y.}~\bibnamefont
  {Peng}}, \bibinfo {author} {\bibfnamefont {Y.}~\bibnamefont {Liu}}, \bibinfo
  {author} {\bibfnamefont {Ch.}\ \bibnamefont {Chen}}, \bibinfo {author}
  {\bibfnamefont {L.}~\bibnamefont {Yu}}, \bibinfo {author} {\bibfnamefont
  {G.}~\bibnamefont {Liu}}, \bibinfo {author} {\bibfnamefont {X.}~\bibnamefont
  {Dong}}, \bibinfo {author} {\bibfnamefont {J.}~\bibnamefont {Zhang}},
  \bibinfo {author} {\bibfnamefont {Ch.}\ \bibnamefont {Chen}}, \bibinfo
  {author} {\bibfnamefont {Z.}~\bibnamefont {Xu}}, \bibinfo {author}
  {\bibfnamefont {X.}~\bibnamefont {Chen}}, \bibinfo {author} {\bibfnamefont
  {X.}~\bibnamefont {Ma}}, \bibinfo {author} {\bibfnamefont {Q.}~\bibnamefont
  {Xue}}, \ and\ \bibinfo {author} {\bibfnamefont {X.~J.}\ \bibnamefont
  {Zhou}},\ }\bibfield  {title} {\enquote {\bibinfo {title} {Phase diagram and
  electronic indication of high-temperature superconductivity at 65 {K} in
  single-layer {FeSe} films},}\ }\href {\doibase 10.1038/nmat3648} {\bibfield
  {journal} {\bibinfo  {journal} {Nat. Mater.}\ }\textbf {\bibinfo {volume}
  {12}},\ \bibinfo {pages} {605} (\bibinfo {year} {2013})}\BibitemShut
  {NoStop}%
\bibitem [{\citenamefont {Tan}\ \emph {et~al.}(2013)\citenamefont {Tan},
  \citenamefont {Zhang}, \citenamefont {Xia}, \citenamefont {Ye}, \citenamefont
  {Chen}, \citenamefont {Xie}, \citenamefont {Peng}, \citenamefont {Xu},
  \citenamefont {Fan}, \citenamefont {Xu}, \citenamefont {Jiang}, \citenamefont
  {Zhang}, \citenamefont {Lai}, \citenamefont {Xiang}, \citenamefont {Hu},
  \citenamefont {Xie},\ and\ \citenamefont {Feng}}]{tan.zhang.13}%
  \BibitemOpen
  \bibfield  {author} {\bibinfo {author} {\bibfnamefont {S.}~\bibnamefont
  {Tan}}, \bibinfo {author} {\bibfnamefont {Y.}~\bibnamefont {Zhang}}, \bibinfo
  {author} {\bibfnamefont {M.}~\bibnamefont {Xia}}, \bibinfo {author}
  {\bibfnamefont {Z.}~\bibnamefont {Ye}}, \bibinfo {author} {\bibfnamefont
  {F.}~\bibnamefont {Chen}}, \bibinfo {author} {\bibfnamefont {X.}~\bibnamefont
  {Xie}}, \bibinfo {author} {\bibfnamefont {R.}~\bibnamefont {Peng}}, \bibinfo
  {author} {\bibfnamefont {D.}~\bibnamefont {Xu}}, \bibinfo {author}
  {\bibfnamefont {Q.}~\bibnamefont {Fan}}, \bibinfo {author} {\bibfnamefont
  {H.}~\bibnamefont {Xu}}, \bibinfo {author} {\bibfnamefont {J.}~\bibnamefont
  {Jiang}}, \bibinfo {author} {\bibfnamefont {T.}~\bibnamefont {Zhang}},
  \bibinfo {author} {\bibfnamefont {X.}~\bibnamefont {Lai}}, \bibinfo {author}
  {\bibfnamefont {T.}~\bibnamefont {Xiang}}, \bibinfo {author} {\bibfnamefont
  {J.}~\bibnamefont {Hu}}, \bibinfo {author} {\bibfnamefont {B.}~\bibnamefont
  {Xie}}, \ and\ \bibinfo {author} {\bibfnamefont {D.}~\bibnamefont {Feng}},\
  }\bibfield  {title} {\enquote {\bibinfo {title} {Interface-induced
  superconductivity and strain-dependent spin density waves in
  {FeSe/SrTiO$_{3}$} thin films},}\ }\href {\doibase 10.1038/nmat3654}
  {\bibfield  {journal} {\bibinfo  {journal} {Nat. Mater.}\ }\textbf {\bibinfo
  {volume} {12}},\ \bibinfo {pages} {634} (\bibinfo {year} {2013})}\BibitemShut
  {NoStop}%
\bibitem [{\citenamefont {Zhou}\ \emph {et~al.}(2016)\citenamefont {Zhou},
  \citenamefont {Zhang}, \citenamefont {Liu}, \citenamefont {Tang},
  \citenamefont {Wang}, \citenamefont {Li}, \citenamefont {Song}, \citenamefont
  {Ji}, \citenamefont {He}, \citenamefont {Wang}, \citenamefont {Ma},\ and\
  \citenamefont {Xue}}]{zhou.zhang.16}%
  \BibitemOpen
  \bibfield  {author} {\bibinfo {author} {\bibfnamefont {G.}~\bibnamefont
  {Zhou}}, \bibinfo {author} {\bibfnamefont {D.}~\bibnamefont {Zhang}},
  \bibinfo {author} {\bibfnamefont {Ch.}\ \bibnamefont {Liu}}, \bibinfo
  {author} {\bibfnamefont {Ch.}\ \bibnamefont {Tang}}, \bibinfo {author}
  {\bibfnamefont {X.}~\bibnamefont {Wang}}, \bibinfo {author} {\bibfnamefont
  {Z.}~\bibnamefont {Li}}, \bibinfo {author} {\bibfnamefont {C.}~\bibnamefont
  {Song}}, \bibinfo {author} {\bibfnamefont {S.}~\bibnamefont {Ji}}, \bibinfo
  {author} {\bibfnamefont {K.}~\bibnamefont {He}}, \bibinfo {author}
  {\bibfnamefont {L.}~\bibnamefont {Wang}}, \bibinfo {author} {\bibfnamefont
  {X.}~\bibnamefont {Ma}}, \ and\ \bibinfo {author} {\bibfnamefont {Q.-K.}\
  \bibnamefont {Xue}},\ }\bibfield  {title} {\enquote {\bibinfo {title}
  {Interface induced high temperature superconductivity in single unit-cell
  {FeSe} on {SrTiO$_{3}$}(110)},}\ }\href {\doibase 10.1063/1.4950964}
  {\bibfield  {journal} {\bibinfo  {journal} {Appl. Phys. Lett.}\ }\textbf
  {\bibinfo {volume} {108}},\ \bibinfo {pages} {202603} (\bibinfo {year}
  {2016})}\BibitemShut {NoStop}%
\bibitem [{\citenamefont {Zhang}\ \emph {et~al.}(2016)\citenamefont {Zhang},
  \citenamefont {Peng}, \citenamefont {Qian}, \citenamefont {Richard},
  \citenamefont {Shi}, \citenamefont {Ma}, \citenamefont {Fu}, \citenamefont
  {Guo}, \citenamefont {Han}, \citenamefont {Wang}, \citenamefont {Wang},
  \citenamefont {Xue}, \citenamefont {Hu}, \citenamefont {Sun},\ and\
  \citenamefont {Ding}}]{zhang.peng.16}%
  \BibitemOpen
  \bibfield  {author} {\bibinfo {author} {\bibfnamefont {P.}~\bibnamefont
  {Zhang}}, \bibinfo {author} {\bibfnamefont {X.-L.}\ \bibnamefont {Peng}},
  \bibinfo {author} {\bibfnamefont {T.}~\bibnamefont {Qian}}, \bibinfo {author}
  {\bibfnamefont {P.}~\bibnamefont {Richard}}, \bibinfo {author} {\bibfnamefont
  {X.}~\bibnamefont {Shi}}, \bibinfo {author} {\bibfnamefont {J.-Z.}\
  \bibnamefont {Ma}}, \bibinfo {author} {\bibfnamefont {B.~B.}\ \bibnamefont
  {Fu}}, \bibinfo {author} {\bibfnamefont {Y.-L.}\ \bibnamefont {Guo}},
  \bibinfo {author} {\bibfnamefont {Z.~Q.}\ \bibnamefont {Han}}, \bibinfo
  {author} {\bibfnamefont {S.~C.}\ \bibnamefont {Wang}}, \bibinfo {author}
  {\bibfnamefont {L.~L.}\ \bibnamefont {Wang}}, \bibinfo {author}
  {\bibfnamefont {Q.-K.}\ \bibnamefont {Xue}}, \bibinfo {author} {\bibfnamefont
  {J.~P.}\ \bibnamefont {Hu}}, \bibinfo {author} {\bibfnamefont {Y.-J.}\
  \bibnamefont {Sun}}, \ and\ \bibinfo {author} {\bibfnamefont
  {H.}~\bibnamefont {Ding}},\ }\bibfield  {title} {\enquote {\bibinfo {title}
  {Observation of high-${T}_{c}$ superconductivity in rectangular
  {$\text{FeSe}/{\mathrm{SrTiO}}_{3}(110)$} monolayers},}\ }\href {\doibase
  10.1103/PhysRevB.94.104510} {\bibfield  {journal} {\bibinfo  {journal} {Phys.
  Rev. B}\ }\textbf {\bibinfo {volume} {94}},\ \bibinfo {pages} {104510}
  (\bibinfo {year} {2016})}\BibitemShut {NoStop}%
\bibitem [{\citenamefont {Wu}\ \emph {et~al.}(2016)\citenamefont {Wu},
  \citenamefont {Dai}, \citenamefont {Liang}, \citenamefont {Le}, \citenamefont
  {Fan},\ and\ \citenamefont {Hu}}]{wu.dai.16}%
  \BibitemOpen
  \bibfield  {author} {\bibinfo {author} {\bibfnamefont {X.}~\bibnamefont
  {Wu}}, \bibinfo {author} {\bibfnamefont {X.}~\bibnamefont {Dai}}, \bibinfo
  {author} {\bibfnamefont {Y.}~\bibnamefont {Liang}}, \bibinfo {author}
  {\bibfnamefont {C.}~\bibnamefont {Le}}, \bibinfo {author} {\bibfnamefont
  {H.}~\bibnamefont {Fan}}, \ and\ \bibinfo {author} {\bibfnamefont
  {J.}~\bibnamefont {Hu}},\ }\bibfield  {title} {\enquote {\bibinfo {title}
  {Density functional calculations of a staggered {FeSe} monolayer on a
  {SrTiO$_{3}$} (110) surface},}\ }\href {\doibase 10.1103/PhysRevB.94.045114}
  {\bibfield  {journal} {\bibinfo  {journal} {Phys. Rev. B}\ }\textbf {\bibinfo
  {volume} {94}},\ \bibinfo {pages} {045114} (\bibinfo {year}
  {2016})}\BibitemShut {NoStop}%
\bibitem [{\citenamefont {Zhong}\ \emph {et~al.}(2013)\citenamefont {Zhong},
  \citenamefont {T\'{o}th},\ and\ \citenamefont {Held}}]{zhong.toth.13}%
  \BibitemOpen
  \bibfield  {author} {\bibinfo {author} {\bibfnamefont {Z.}~\bibnamefont
  {Zhong}}, \bibinfo {author} {\bibfnamefont {A.}~\bibnamefont {T\'{o}th}}, \
  and\ \bibinfo {author} {\bibfnamefont {K.}~\bibnamefont {Held}},\ }\bibfield
  {title} {\enquote {\bibinfo {title} {Theory of spin-orbit coupling at
  {LaAlO$_{3}$/SrTiO$_{3}$} interfaces and {SrTiO$_{3}$} surfaces},}\ }\href
  {\doibase 10.1103/PhysRevB.87.161102} {\bibfield  {journal} {\bibinfo
  {journal} {Phys. Rev. B}\ }\textbf {\bibinfo {volume} {87}},\ \bibinfo
  {pages} {161102} (\bibinfo {year} {2013})}\BibitemShut {NoStop}%
\bibitem [{\citenamefont {Wilson}\ \emph {et~al.}(2017)\citenamefont {Wilson},
  \citenamefont {Nguyen}, \citenamefont {Seyler}, \citenamefont {Rivera},
  \citenamefont {Marsden}, \citenamefont {Laker}, \citenamefont
  {Constantinescu}, \citenamefont {Kandyba}, \citenamefont {Barinov},
  \citenamefont {Hine}, \citenamefont {Xu},\ and\ \citenamefont
  {Cobden}}]{wilson.nguyen.17}%
  \BibitemOpen
  \bibfield  {author} {\bibinfo {author} {\bibfnamefont {N.~R.}\ \bibnamefont
  {Wilson}}, \bibinfo {author} {\bibfnamefont {P.~V.}\ \bibnamefont {Nguyen}},
  \bibinfo {author} {\bibfnamefont {K.}~\bibnamefont {Seyler}}, \bibinfo
  {author} {\bibfnamefont {P.}~\bibnamefont {Rivera}}, \bibinfo {author}
  {\bibfnamefont {A.~J.}\ \bibnamefont {Marsden}}, \bibinfo {author}
  {\bibfnamefont {Z.~P.~L.}\ \bibnamefont {Laker}}, \bibinfo {author}
  {\bibfnamefont {G.~C.}\ \bibnamefont {Constantinescu}}, \bibinfo {author}
  {\bibfnamefont {V.}~\bibnamefont {Kandyba}}, \bibinfo {author} {\bibfnamefont
  {A.}~\bibnamefont {Barinov}}, \bibinfo {author} {\bibfnamefont {N.~D.~M.}\
  \bibnamefont {Hine}}, \bibinfo {author} {\bibfnamefont {X.}~\bibnamefont
  {Xu}}, \ and\ \bibinfo {author} {\bibfnamefont {D.~H.}\ \bibnamefont
  {Cobden}},\ }\bibfield  {title} {\enquote {\bibinfo {title} {Determination of
  band offsets, hybridization, and exciton binding in {2D} semiconductor
  heterostructures},}\ }\href {\doibase 10.1126/sciadv.1601832} {\bibfield
  {journal} {\bibinfo  {journal} {Science Advances}\ }\textbf {\bibinfo
  {volume} {3}},\ \bibinfo {pages} {e1601832} (\bibinfo {year}
  {2017})}\BibitemShut {NoStop}%
\bibitem [{\citenamefont {He}\ \emph {et~al.}(2016)\citenamefont {He},
  \citenamefont {Shen}, \citenamefont {Petrovi\'{c}}, \citenamefont {He},
  \citenamefont {Liu}, \citenamefont {Zheng}, \citenamefont {Wong},
  \citenamefont {Chen}, \citenamefont {Wang}, \citenamefont {Law},
  \citenamefont {Sou},\ and\ \citenamefont {Lortz}}]{he.shen.16}%
  \BibitemOpen
  \bibfield  {author} {\bibinfo {author} {\bibfnamefont {M.~Q.}\ \bibnamefont
  {He}}, \bibinfo {author} {\bibfnamefont {J.~Y.}\ \bibnamefont {Shen}},
  \bibinfo {author} {\bibfnamefont {A.~P.}\ \bibnamefont {Petrovi\'{c}}},
  \bibinfo {author} {\bibfnamefont {Q.~L.}\ \bibnamefont {He}}, \bibinfo
  {author} {\bibfnamefont {H.~C.}\ \bibnamefont {Liu}}, \bibinfo {author}
  {\bibfnamefont {Y.}~\bibnamefont {Zheng}}, \bibinfo {author} {\bibfnamefont
  {C.~H.}\ \bibnamefont {Wong}}, \bibinfo {author} {\bibfnamefont {Q.~H.}\
  \bibnamefont {Chen}}, \bibinfo {author} {\bibfnamefont {J.~N.}\ \bibnamefont
  {Wang}}, \bibinfo {author} {\bibfnamefont {K.~T.}\ \bibnamefont {Law}},
  \bibinfo {author} {\bibfnamefont {I.~K.}\ \bibnamefont {Sou}}, \ and\
  \bibinfo {author} {\bibfnamefont {R.}~\bibnamefont {Lortz}},\ }\bibfield
  {title} {\enquote {\bibinfo {title} {Pseudogap and proximity effect in the
  {Bi$_{2}$Te$_{3}$/Fe$_{1+y}$Te} interfacial superconductor},}\ }\href
  {\doibase 10.1038/srep32508} {\bibfield  {journal} {\bibinfo  {journal} {Sci.
  Rep.}\ }\textbf {\bibinfo {volume} {6}},\ \bibinfo {pages} {32508} (\bibinfo
  {year} {2016})}\BibitemShut {NoStop}%
\bibitem [{\citenamefont {Avsar}\ \emph {et~al.}(2014)\citenamefont {Avsar},
  \citenamefont {Tan}, \citenamefont {Taychatanapat}, \citenamefont
  {Balakrishnan}, \citenamefont {Koon}, \citenamefont {Yeo}, \citenamefont
  {Lahiri}, \citenamefont {Carvalho}, \citenamefont {Rodin}, \citenamefont
  {O'Farrell}, \citenamefont {Eda}, \citenamefont {Castro~Neto},\ and\
  \citenamefont {\"{O}zyilmaz}}]{avsar.tan.14}%
  \BibitemOpen
  \bibfield  {author} {\bibinfo {author} {\bibfnamefont {A.}~\bibnamefont
  {Avsar}}, \bibinfo {author} {\bibfnamefont {J.~Y.}\ \bibnamefont {Tan}},
  \bibinfo {author} {\bibfnamefont {T.}~\bibnamefont {Taychatanapat}}, \bibinfo
  {author} {\bibfnamefont {J.}~\bibnamefont {Balakrishnan}}, \bibinfo {author}
  {\bibfnamefont {G.~K.~W.}\ \bibnamefont {Koon}}, \bibinfo {author}
  {\bibfnamefont {Y.}~\bibnamefont {Yeo}}, \bibinfo {author} {\bibfnamefont
  {J.}~\bibnamefont {Lahiri}}, \bibinfo {author} {\bibfnamefont
  {A.}~\bibnamefont {Carvalho}}, \bibinfo {author} {\bibfnamefont {A.~S.}\
  \bibnamefont {Rodin}}, \bibinfo {author} {\bibfnamefont {E.~C.~T.}\
  \bibnamefont {O'Farrell}}, \bibinfo {author} {\bibfnamefont {G.}~\bibnamefont
  {Eda}}, \bibinfo {author} {\bibfnamefont {A.~H.}\ \bibnamefont
  {Castro~Neto}}, \ and\ \bibinfo {author} {\bibfnamefont {B.}~\bibnamefont
  {\"{O}zyilmaz}},\ }\bibfield  {title} {\enquote {\bibinfo {title}
  {Spin--orbit proximity effect in graphene},}\ }\href {\doibase
  10.1038/ncomms5875} {\bibfield  {journal} {\bibinfo  {journal} {Nat.
  Commun.}\ }\textbf {\bibinfo {volume} {5}},\ \bibinfo {pages} {4875}
  (\bibinfo {year} {2014})}\BibitemShut {NoStop}%
\bibitem [{\citenamefont {Marchenko}\ \emph {et~al.}(2012)\citenamefont
  {Marchenko}, \citenamefont {Varykhalov}, \citenamefont {Scholz},
  \citenamefont {Bihlmayer}, \citenamefont {Rashba}, \citenamefont {Rybkin},
  \citenamefont {Shikin},\ and\ \citenamefont
  {Rader}}]{marchenko.varykhalov.12}%
  \BibitemOpen
  \bibfield  {author} {\bibinfo {author} {\bibfnamefont {D.}~\bibnamefont
  {Marchenko}}, \bibinfo {author} {\bibfnamefont {A.}~\bibnamefont
  {Varykhalov}}, \bibinfo {author} {\bibfnamefont {M.~R.}\ \bibnamefont
  {Scholz}}, \bibinfo {author} {\bibfnamefont {G.}~\bibnamefont {Bihlmayer}},
  \bibinfo {author} {\bibfnamefont {E.~I.}\ \bibnamefont {Rashba}}, \bibinfo
  {author} {\bibfnamefont {A.}~\bibnamefont {Rybkin}}, \bibinfo {author}
  {\bibfnamefont {A.~M.}\ \bibnamefont {Shikin}}, \ and\ \bibinfo {author}
  {\bibfnamefont {O.}~\bibnamefont {Rader}},\ }\bibfield  {title} {\enquote
  {\bibinfo {title} {Giant {Rashba} splitting in graphene due to hybridization
  with gold},}\ }\href {http://doi.org/10.1038/ncomms2227} {\bibfield
  {journal} {\bibinfo  {journal} {Nat. Commun.}\ }\textbf {\bibinfo {volume}
  {3}},\ \bibinfo {pages} {1232} (\bibinfo {year} {2012})}\BibitemShut
  {NoStop}%
\bibitem [{\citenamefont {Qiao}\ \emph {et~al.}(2014)\citenamefont {Qiao},
  \citenamefont {Ren}, \citenamefont {Chen}, \citenamefont {Bellaiche},
  \citenamefont {Zhang}, \citenamefont {MacDonald},\ and\ \citenamefont
  {Niu}}]{qiao.ren.14}%
  \BibitemOpen
  \bibfield  {author} {\bibinfo {author} {\bibfnamefont {Z.}~\bibnamefont
  {Qiao}}, \bibinfo {author} {\bibfnamefont {W.}~\bibnamefont {Ren}}, \bibinfo
  {author} {\bibfnamefont {H.}~\bibnamefont {Chen}}, \bibinfo {author}
  {\bibfnamefont {L.}~\bibnamefont {Bellaiche}}, \bibinfo {author}
  {\bibfnamefont {Z.}~\bibnamefont {Zhang}}, \bibinfo {author} {\bibfnamefont
  {A.~H.}\ \bibnamefont {MacDonald}}, \ and\ \bibinfo {author} {\bibfnamefont
  {Q.}~\bibnamefont {Niu}},\ }\bibfield  {title} {\enquote {\bibinfo {title}
  {Quantum anomalous hall effect in graphene proximity coupled to an
  antiferromagnetic insulator},}\ }\href {\doibase
  10.1103/PhysRevLett.112.116404} {\bibfield  {journal} {\bibinfo  {journal}
  {Phys. Rev. Lett.}\ }\textbf {\bibinfo {volume} {112}},\ \bibinfo {pages}
  {116404} (\bibinfo {year} {2014})}\BibitemShut {NoStop}%
\bibitem [{\citenamefont {Chudzinski}(2015)}]{chudzinski.15}%
  \BibitemOpen
  \bibfield  {author} {\bibinfo {author} {\bibfnamefont {P.}~\bibnamefont
  {Chudzinski}},\ }\bibfield  {title} {\enquote {\bibinfo {title} {Spin-orbit
  coupling and proximity effects in metallic carbon nanotubes},}\ }\href
  {\doibase 10.1103/PhysRevB.92.115147} {\bibfield  {journal} {\bibinfo
  {journal} {Phys. Rev. B}\ }\textbf {\bibinfo {volume} {92}},\ \bibinfo
  {pages} {115147} (\bibinfo {year} {2015})}\BibitemShut {NoStop}%
\bibitem [{\citenamefont {Zhang}\ \emph {et~al.}(2015)\citenamefont {Zhang},
  \citenamefont {Sun}, \citenamefont {Che}, \citenamefont {Li}, \citenamefont
  {Zhang}, \citenamefont {Shan}, \citenamefont {Zhu},\ and\ \citenamefont
  {Su}}]{zhang.sun.15}%
  \BibitemOpen
  \bibfield  {author} {\bibinfo {author} {\bibfnamefont {Y.~Q.}\ \bibnamefont
  {Zhang}}, \bibinfo {author} {\bibfnamefont {N.~Y.}\ \bibnamefont {Sun}},
  \bibinfo {author} {\bibfnamefont {W.~R.}\ \bibnamefont {Che}}, \bibinfo
  {author} {\bibfnamefont {X.~L.}\ \bibnamefont {Li}}, \bibinfo {author}
  {\bibfnamefont {J.~W.}\ \bibnamefont {Zhang}}, \bibinfo {author}
  {\bibfnamefont {R.}~\bibnamefont {Shan}}, \bibinfo {author} {\bibfnamefont
  {Z.~G.}\ \bibnamefont {Zhu}}, \ and\ \bibinfo {author} {\bibfnamefont
  {G.}~\bibnamefont {Su}},\ }\bibfield  {title} {\enquote {\bibinfo {title}
  {Manipulating effective spin orbit coupling based on proximity effect in
  magnetic bilayers},}\ }\href {\doibase 10.1063/1.4929585} {\bibfield
  {journal} {\bibinfo  {journal} {Appl. Phys. Lett.}\ }\textbf {\bibinfo
  {volume} {107}},\ \bibinfo {pages} {082404} (\bibinfo {year}
  {2015})}\BibitemShut {NoStop}%
\bibitem [{\citenamefont {Balakrishnan}\ \emph {et~al.}(2013)\citenamefont
  {Balakrishnan}, \citenamefont {Kok Wai~Koon}, \citenamefont {Jaiswal},
  \citenamefont {Castro~N.},\ and\ \citenamefont
  {Ozyilmaz}}]{balakrishnan.kokwaikoon.13}%
  \BibitemOpen
  \bibfield  {author} {\bibinfo {author} {\bibfnamefont {J.}~\bibnamefont
  {Balakrishnan}}, \bibinfo {author} {\bibfnamefont {G.}~\bibnamefont {Kok
  Wai~Koon}}, \bibinfo {author} {\bibfnamefont {M.}~\bibnamefont {Jaiswal}},
  \bibinfo {author} {\bibfnamefont {A.~H.}\ \bibnamefont {Castro~N.}}, \ and\
  \bibinfo {author} {\bibfnamefont {B.}~\bibnamefont {Ozyilmaz}},\ }\bibfield
  {title} {\enquote {\bibinfo {title} {Colossal enhancement of spin-orbit
  coupling in weakly hydrogenated graphene},}\ }\href {\doibase
  10.1038/nphys2576} {\bibfield  {journal} {\bibinfo  {journal} {Nat. Phys.}\
  }\textbf {\bibinfo {volume} {9}},\ \bibinfo {pages} {284} (\bibinfo {year}
  {2013})}\BibitemShut {NoStop}%
\bibitem [{\citenamefont {Castro~Neto}\ and\ \citenamefont
  {Guinea}(2009)}]{castroneto.guinea.09}%
  \BibitemOpen
  \bibfield  {author} {\bibinfo {author} {\bibfnamefont {A.~H.}\ \bibnamefont
  {Castro~Neto}}\ and\ \bibinfo {author} {\bibfnamefont {F.}~\bibnamefont
  {Guinea}},\ }\bibfield  {title} {\enquote {\bibinfo {title} {Impurity-induced
  spin-orbit coupling in graphene},}\ }\href {\doibase
  10.1103/PhysRevLett.103.026804} {\bibfield  {journal} {\bibinfo  {journal}
  {Phys. Rev. Lett.}\ }\textbf {\bibinfo {volume} {103}},\ \bibinfo {pages}
  {026804} (\bibinfo {year} {2009})}\BibitemShut {NoStop}%
\bibitem [{\citenamefont {Fu}\ \emph {et~al.}(2007)\citenamefont {Fu},
  \citenamefont {Kane},\ and\ \citenamefont {Mele}}]{fu.kane.07}%
  \BibitemOpen
  \bibfield  {author} {\bibinfo {author} {\bibfnamefont {L.}~\bibnamefont
  {Fu}}, \bibinfo {author} {\bibfnamefont {C.~L.}\ \bibnamefont {Kane}}, \ and\
  \bibinfo {author} {\bibfnamefont {E.~J.}\ \bibnamefont {Mele}},\ }\bibfield
  {title} {\enquote {\bibinfo {title} {Topological insulators in three
  dimensions},}\ }\href {\doibase 10.1103/PhysRevLett.98.106803} {\bibfield
  {journal} {\bibinfo  {journal} {Phys. Rev. Lett.}\ }\textbf {\bibinfo
  {volume} {98}},\ \bibinfo {pages} {106803} (\bibinfo {year}
  {2007})}\BibitemShut {NoStop}%
\bibitem [{\citenamefont {Qi}\ and\ \citenamefont {Zhang}(2011)}]{qi.zhang.11}%
  \BibitemOpen
  \bibfield  {author} {\bibinfo {author} {\bibfnamefont {X.-L.}\ \bibnamefont
  {Qi}}\ and\ \bibinfo {author} {\bibfnamefont {S.-Ch.}\ \bibnamefont
  {Zhang}},\ }\bibfield  {title} {\enquote {\bibinfo {title} {Topological
  insulators and superconductors},}\ }\href {\doibase
  10.1103/RevModPhys.83.1057} {\bibfield  {journal} {\bibinfo  {journal} {Rev.
  Mod. Phys.}\ }\textbf {\bibinfo {volume} {83}},\ \bibinfo {pages} {1057}
  (\bibinfo {year} {2011})}\BibitemShut {NoStop}%
\bibitem [{\citenamefont {Mackenzie}\ and\ \citenamefont
  {Maeno}(2003)}]{mackenzie.maeno.03}%
  \BibitemOpen
  \bibfield  {author} {\bibinfo {author} {\bibfnamefont {A.~P.}\ \bibnamefont
  {Mackenzie}}\ and\ \bibinfo {author} {\bibfnamefont {Y.}~\bibnamefont
  {Maeno}},\ }\bibfield  {title} {\enquote {\bibinfo {title} {The
  superconductivity of {Sr$_{2}$RuO$_{4}$} and the physics of spin-triplet
  pairing},}\ }\href {\doibase 10.1103/RevModPhys.75.657} {\bibfield  {journal}
  {\bibinfo  {journal} {Rev. Mod. Phys.}\ }\textbf {\bibinfo {volume} {75}},\
  \bibinfo {pages} {657} (\bibinfo {year} {2003})}\BibitemShut {NoStop}%
\bibitem [{\citenamefont {Bergeret}\ \emph {et~al.}(2005)\citenamefont
  {Bergeret}, \citenamefont {Volkov},\ and\ \citenamefont
  {Efetov}}]{bergeret.volkov.05}%
  \BibitemOpen
  \bibfield  {author} {\bibinfo {author} {\bibfnamefont {F.~S.}\ \bibnamefont
  {Bergeret}}, \bibinfo {author} {\bibfnamefont {A.~F.}\ \bibnamefont
  {Volkov}}, \ and\ \bibinfo {author} {\bibfnamefont {K.~B.}\ \bibnamefont
  {Efetov}},\ }\bibfield  {title} {\enquote {\bibinfo {title} {Odd triplet
  superconductivity and related phenomena in superconductor-ferromagnet
  structures},}\ }\href {\doibase 10.1103/RevModPhys.77.1321} {\bibfield
  {journal} {\bibinfo  {journal} {Rev. Mod. Phys.}\ }\textbf {\bibinfo {volume}
  {77}},\ \bibinfo {pages} {1321} (\bibinfo {year} {2005})}\BibitemShut
  {NoStop}%
\bibitem [{\citenamefont {Sato}\ and\ \citenamefont
  {Fujimoto}(2009)}]{sato.fujimoto.09}%
  \BibitemOpen
  \bibfield  {author} {\bibinfo {author} {\bibfnamefont {M.}~\bibnamefont
  {Sato}}\ and\ \bibinfo {author} {\bibfnamefont {S.}~\bibnamefont
  {Fujimoto}},\ }\bibfield  {title} {\enquote {\bibinfo {title} {Topological
  phases of noncentrosymmetric superconductors: Edge states, {Majorana}
  fermions, and non-{Abelian} statistics},}\ }\href {\doibase
  10.1103/PhysRevB.79.094504} {\bibfield  {journal} {\bibinfo  {journal} {Phys.
  Rev. B}\ }\textbf {\bibinfo {volume} {79}},\ \bibinfo {pages} {094504}
  (\bibinfo {year} {2009})}\BibitemShut {NoStop}%
\bibitem [{\citenamefont {Smidman}\ \emph {et~al.}(2017)\citenamefont
  {Smidman}, \citenamefont {Salamon}, \citenamefont {Yuan},\ and\ \citenamefont
  {Agterberg}}]{smidman.salamon.17}%
  \BibitemOpen
  \bibfield  {author} {\bibinfo {author} {\bibfnamefont {M.}~\bibnamefont
  {Smidman}}, \bibinfo {author} {\bibfnamefont {M.~B.}\ \bibnamefont
  {Salamon}}, \bibinfo {author} {\bibfnamefont {H.~Q.}\ \bibnamefont {Yuan}}, \
  and\ \bibinfo {author} {\bibfnamefont {D.~F.}\ \bibnamefont {Agterberg}},\
  }\bibfield  {title} {\enquote {\bibinfo {title} {Superconductivity and
  spin-orbit coupling in non-centrosymmetric materials: a review},}\ }\href
  {\doibase 10.1088/1361-6633/80/3/036501} {\bibfield  {journal} {\bibinfo
  {journal} {Rep. Prog. Phys.}\ }\textbf {\bibinfo {volume} {80}},\ \bibinfo
  {pages} {036501} (\bibinfo {year} {2017})}\BibitemShut {NoStop}%
\bibitem [{\citenamefont {Sato}\ \emph {et~al.}(2010)\citenamefont {Sato},
  \citenamefont {Takahashi},\ and\ \citenamefont
  {Fujimoto}}]{sato.takahasi.10}%
  \BibitemOpen
  \bibfield  {author} {\bibinfo {author} {\bibfnamefont {M.}~\bibnamefont
  {Sato}}, \bibinfo {author} {\bibfnamefont {Y.}~\bibnamefont {Takahashi}}, \
  and\ \bibinfo {author} {\bibfnamefont {S.}~\bibnamefont {Fujimoto}},\
  }\bibfield  {title} {\enquote {\bibinfo {title} {{Non-Abelian} topological
  orders and majorana fermions in spin-singlet superconductors},}\ }\href
  {\doibase 10.1103/PhysRevB.82.134521} {\bibfield  {journal} {\bibinfo
  {journal} {Phys. Rev. B}\ }\textbf {\bibinfo {volume} {82}},\ \bibinfo
  {pages} {134521} (\bibinfo {year} {2010})}\BibitemShut {NoStop}%
\bibitem [{\citenamefont {Gor'kov}\ and\ \citenamefont
  {Rashba}(2001)}]{gorkov.rashba.01}%
  \BibitemOpen
  \bibfield  {author} {\bibinfo {author} {\bibfnamefont {L.~P.}\ \bibnamefont
  {Gor'kov}}\ and\ \bibinfo {author} {\bibfnamefont {E.~I.}\ \bibnamefont
  {Rashba}},\ }\bibfield  {title} {\enquote {\bibinfo {title} {Superconducting
  {2D} system with lifted spin degeneracy: Mixed singlet-triplet state},}\
  }\href {\doibase 10.1103/PhysRevLett.87.037004} {\bibfield  {journal}
  {\bibinfo  {journal} {Phys. Rev. Lett.}\ }\textbf {\bibinfo {volume} {87}},\
  \bibinfo {pages} {037004} (\bibinfo {year} {2001})}\BibitemShut {NoStop}%
\bibitem [{\citenamefont {Zhang}\ \emph {et~al.}(2008)\citenamefont {Zhang},
  \citenamefont {Tewari}, \citenamefont {Lutchyn},\ and\ \citenamefont
  {Das~Sarma}}]{zhang.tewari.08}%
  \BibitemOpen
  \bibfield  {author} {\bibinfo {author} {\bibfnamefont {Ch.}\ \bibnamefont
  {Zhang}}, \bibinfo {author} {\bibfnamefont {S.}~\bibnamefont {Tewari}},
  \bibinfo {author} {\bibfnamefont {Roman~M.}\ \bibnamefont {Lutchyn}}, \ and\
  \bibinfo {author} {\bibfnamefont {S.}~\bibnamefont {Das~Sarma}},\ }\bibfield
  {title} {\enquote {\bibinfo {title} {${p}_{x}+i{p}_{y}$ superfluid from {\it
  s}-wave interactions of fermionic cold atoms},}\ }\href {\doibase
  10.1103/PhysRevLett.101.160401} {\bibfield  {journal} {\bibinfo  {journal}
  {Phys. Rev. Lett.}\ }\textbf {\bibinfo {volume} {101}},\ \bibinfo {pages}
  {160401} (\bibinfo {year} {2008})}\BibitemShut {NoStop}%
\bibitem [{\citenamefont {Alicea}(2010)}]{alicae.10}%
  \BibitemOpen
  \bibfield  {author} {\bibinfo {author} {\bibfnamefont {J.}~\bibnamefont
  {Alicea}},\ }\bibfield  {title} {\enquote {\bibinfo {title} {Majorana
  fermions in a tunable semiconductor device},}\ }\href {\doibase
  10.1103/PhysRevB.81.125318} {\bibfield  {journal} {\bibinfo  {journal} {Phys.
  Rev. B}\ }\textbf {\bibinfo {volume} {81}},\ \bibinfo {pages} {125318}
  (\bibinfo {year} {2010})}\BibitemShut {NoStop}%
\bibitem [{\citenamefont {Seo}\ \emph {et~al.}(2012)\citenamefont {Seo},
  \citenamefont {Han},\ and\ \citenamefont {S\'a~de Melo}}]{seo.han.12}%
  \BibitemOpen
  \bibfield  {author} {\bibinfo {author} {\bibfnamefont {K.}~\bibnamefont
  {Seo}}, \bibinfo {author} {\bibfnamefont {L.}~\bibnamefont {Han}}, \ and\
  \bibinfo {author} {\bibfnamefont {C.~A.~R.}\ \bibnamefont {S\'a~de Melo}},\
  }\bibfield  {title} {\enquote {\bibinfo {title} {Topological phase
  transitions in ultracold {Fermi} superfluids: The evolution from
  {Bardeen-Cooper-Schrieffer} to {Bose-Einstein}-condensate superfluids under
  artificial spin-orbit fields},}\ }\href {\doibase 10.1103/PhysRevA.85.033601}
  {\bibfield  {journal} {\bibinfo  {journal} {Phys. Rev. A}\ }\textbf {\bibinfo
  {volume} {85}},\ \bibinfo {pages} {033601} (\bibinfo {year}
  {2012})}\BibitemShut {NoStop}%
\bibitem [{\citenamefont {Yu}\ and\ \citenamefont {Wu}(2016)}]{yu.wu.16}%
  \BibitemOpen
  \bibfield  {author} {\bibinfo {author} {\bibfnamefont {T.}~\bibnamefont
  {Yu}}\ and\ \bibinfo {author} {\bibfnamefont {M.~W.}\ \bibnamefont {Wu}},\
  }\bibfield  {title} {\enquote {\bibinfo {title} {Gapped triplet $p$-wave
  superconductivity in strong spin-orbit-coupled semiconductor quantum wells in
  proximity to $s$-wave superconductor},}\ }\href {\doibase
  10.1103/PhysRevB.93.195308} {\bibfield  {journal} {\bibinfo  {journal} {Phys.
  Rev. B}\ }\textbf {\bibinfo {volume} {93}},\ \bibinfo {pages} {195308}
  (\bibinfo {year} {2016})}\BibitemShut {NoStop}%
\bibitem [{\citenamefont {Tanaka}\ \emph {et~al.}(2009)\citenamefont {Tanaka},
  \citenamefont {Yokoyama},\ and\ \citenamefont
  {Nagaosa}}]{tanaka.yokoyama.09}%
  \BibitemOpen
  \bibfield  {author} {\bibinfo {author} {\bibfnamefont {Y.}~\bibnamefont
  {Tanaka}}, \bibinfo {author} {\bibfnamefont {T.}~\bibnamefont {Yokoyama}}, \
  and\ \bibinfo {author} {\bibfnamefont {N.}~\bibnamefont {Nagaosa}},\
  }\bibfield  {title} {\enquote {\bibinfo {title} {Manipulation of the
  {Majorana} fermion, {Andreev} reflection, and {Josephson} current on
  topological insulators},}\ }\href {\doibase 10.1103/PhysRevLett.103.107002}
  {\bibfield  {journal} {\bibinfo  {journal} {Phys. Rev. Lett.}\ }\textbf
  {\bibinfo {volume} {103}},\ \bibinfo {pages} {107002} (\bibinfo {year}
  {2009})}\BibitemShut {NoStop}%
\bibitem [{\citenamefont {Li}\ \emph {et~al.}(2011)\citenamefont {Li},
  \citenamefont {Covaci}, \citenamefont {Berciu}, \citenamefont {Baillie},\
  and\ \citenamefont {Marsiglio}}]{li.covaci.11}%
  \BibitemOpen
  \bibfield  {author} {\bibinfo {author} {\bibfnamefont {Z.}~\bibnamefont
  {Li}}, \bibinfo {author} {\bibfnamefont {L.}~\bibnamefont {Covaci}}, \bibinfo
  {author} {\bibfnamefont {M.}~\bibnamefont {Berciu}}, \bibinfo {author}
  {\bibfnamefont {D.}~\bibnamefont {Baillie}}, \ and\ \bibinfo {author}
  {\bibfnamefont {F.}~\bibnamefont {Marsiglio}},\ }\bibfield  {title} {\enquote
  {\bibinfo {title} {Impact of spin-orbit coupling on the {Holstein}
  polaron},}\ }\href {\doibase 10.1103/PhysRevB.83.195104} {\bibfield
  {journal} {\bibinfo  {journal} {Phys. Rev. B}\ }\textbf {\bibinfo {volume}
  {83}},\ \bibinfo {pages} {195104} (\bibinfo {year} {2011})}\BibitemShut
  {NoStop}%
\bibitem [{\citenamefont {Li}\ \emph {et~al.}(2012)\citenamefont {Li},
  \citenamefont {Covaci},\ and\ \citenamefont {Marsiglio}}]{li.covaci.12}%
  \BibitemOpen
  \bibfield  {author} {\bibinfo {author} {\bibfnamefont {Z.}~\bibnamefont
  {Li}}, \bibinfo {author} {\bibfnamefont {L.}~\bibnamefont {Covaci}}, \ and\
  \bibinfo {author} {\bibfnamefont {F.}~\bibnamefont {Marsiglio}},\ }\bibfield
  {title} {\enquote {\bibinfo {title} {Impact of {Dresselhaus} versus {Rashba}
  spin-orbit coupling on the {Holstein} polaron},}\ }\href {\doibase
  10.1103/PhysRevB.85.205112} {\bibfield  {journal} {\bibinfo  {journal} {Phys.
  Rev. B}\ }\textbf {\bibinfo {volume} {85}},\ \bibinfo {pages} {205112}
  (\bibinfo {year} {2012})}\BibitemShut {NoStop}%
\bibitem [{\citenamefont {Ptok}\ and\ \citenamefont
  {Crivelli}(2013)}]{ptok.crivelli.13}%
  \BibitemOpen
  \bibfield  {author} {\bibinfo {author} {\bibfnamefont {A.}~\bibnamefont
  {Ptok}}\ and\ \bibinfo {author} {\bibfnamefont {D.}~\bibnamefont
  {Crivelli}},\ }\bibfield  {title} {\enquote {\bibinfo {title} {The
  {Fulde-Ferrell-Larkin-Ovchinnikov} state in pnictides},}\ }\href {\doibase
  10.1007/s10909-013-0871-0} {\bibfield  {journal} {\bibinfo  {journal} {J. Low
  Temp. Phys.}\ }\textbf {\bibinfo {volume} {172}},\ \bibinfo {pages} {226}
  (\bibinfo {year} {2013})}\BibitemShut {NoStop}%
\bibitem [{\citenamefont {Halboth}\ and\ \citenamefont
  {Metzner}(2000)}]{halboth.walter.00}%
  \BibitemOpen
  \bibfield  {author} {\bibinfo {author} {\bibfnamefont {C.~J.}\ \bibnamefont
  {Halboth}}\ and\ \bibinfo {author} {\bibfnamefont {W.}~\bibnamefont
  {Metzner}},\ }\bibfield  {title} {\enquote {\bibinfo {title} {{\it d}-wave
  superconductivity and {Pomeranchuk} instability in the two-dimensional
  {Hubbard} model},}\ }\href {\doibase 10.1103/PhysRevLett.85.5162} {\bibfield
  {journal} {\bibinfo  {journal} {Phys. Rev. Lett.}\ }\textbf {\bibinfo
  {volume} {85}},\ \bibinfo {pages} {5162} (\bibinfo {year}
  {2000})}\BibitemShut {NoStop}%
\bibitem [{\citenamefont {Ptok}\ \emph {et~al.}(2015)\citenamefont {Ptok},
  \citenamefont {Crivelli},\ and\ \citenamefont {Kapcia}}]{ptok.crivelli.15}%
  \BibitemOpen
  \bibfield  {author} {\bibinfo {author} {\bibfnamefont {A.}~\bibnamefont
  {Ptok}}, \bibinfo {author} {\bibfnamefont {D.}~\bibnamefont {Crivelli}}, \
  and\ \bibinfo {author} {\bibfnamefont {K.~J.}\ \bibnamefont {Kapcia}},\
  }\bibfield  {title} {\enquote {\bibinfo {title} {Change of the sign of
  superconducting intraband order parameters induced by interband pair hopping
  interaction in iron-based high-temperature superconductors},}\ }\href
  {\doibase 10.1088/0953-2048/28/4/045010} {\bibfield  {journal} {\bibinfo
  {journal} {Supercond. Sci. Technol.}\ }\textbf {\bibinfo {volume} {28}},\
  \bibinfo {pages} {045010} (\bibinfo {year} {2015})}\BibitemShut {NoStop}%
\bibitem [{\citenamefont {Ptok}(2014)}]{ptok.14}%
  \BibitemOpen
  \bibfield  {author} {\bibinfo {author} {\bibfnamefont {A.}~\bibnamefont
  {Ptok}},\ }\bibfield  {title} {\enquote {\bibinfo {title} {Influence of
  s$_{\pm}$ symmetry on unconventional superconductivity in pnictides above the
  {Pauli} limit -- two-band model study},}\ }\href {\doibase
  10.1140/epjb/e2013-41007-2} {\bibfield  {journal} {\bibinfo  {journal} {Eur.
  Phys. J. B}\ }\textbf {\bibinfo {volume} {87}},\ \bibinfo {pages} {2}
  (\bibinfo {year} {2014})}\BibitemShut {NoStop}%
\bibitem [{\citenamefont {Micnas}\ \emph {et~al.}(1990)\citenamefont {Micnas},
  \citenamefont {Ranninger},\ and\ \citenamefont
  {Robaszkiewicz}}]{micnas.ranninger.90}%
  \BibitemOpen
  \bibfield  {author} {\bibinfo {author} {\bibfnamefont {R.}~\bibnamefont
  {Micnas}}, \bibinfo {author} {\bibfnamefont {J.}~\bibnamefont {Ranninger}}, \
  and\ \bibinfo {author} {\bibfnamefont {S.}~\bibnamefont {Robaszkiewicz}},\
  }\bibfield  {title} {\enquote {\bibinfo {title} {Superconductivity in
  narrow-band systems with local nonretarded attractive interactions},}\ }\href
  {\doibase 10.1103/RevModPhys.62.113} {\bibfield  {journal} {\bibinfo
  {journal} {Rev. Mod. Phys.}\ }\textbf {\bibinfo {volume} {62}},\ \bibinfo
  {pages} {113} (\bibinfo {year} {1990})}\BibitemShut {NoStop}%
\bibitem [{\citenamefont {Bardeen}\ \emph
  {et~al.}(1957{\natexlab{a}})\citenamefont {Bardeen}, \citenamefont {Cooper},\
  and\ \citenamefont {Schrieffer}}]{bardeen.cooper.57a}%
  \BibitemOpen
  \bibfield  {author} {\bibinfo {author} {\bibfnamefont {J.}~\bibnamefont
  {Bardeen}}, \bibinfo {author} {\bibfnamefont {L.~N.}\ \bibnamefont {Cooper}},
  \ and\ \bibinfo {author} {\bibfnamefont {J.~R.}\ \bibnamefont {Schrieffer}},\
  }\bibfield  {title} {\enquote {\bibinfo {title} {Microscopic theory of
  superconductivity},}\ }\href {\doibase 10.1103/PhysRev.106.162} {\bibfield
  {journal} {\bibinfo  {journal} {Phys. Rev.}\ }\textbf {\bibinfo {volume}
  {106}},\ \bibinfo {pages} {162} (\bibinfo {year}
  {1957}{\natexlab{a}})}\BibitemShut {NoStop}%
\bibitem [{\citenamefont {Bardeen}\ \emph
  {et~al.}(1957{\natexlab{b}})\citenamefont {Bardeen}, \citenamefont {Cooper},\
  and\ \citenamefont {Schrieffer}}]{bardeen.cooper.57b}%
  \BibitemOpen
  \bibfield  {author} {\bibinfo {author} {\bibfnamefont {J.}~\bibnamefont
  {Bardeen}}, \bibinfo {author} {\bibfnamefont {L.~N.}\ \bibnamefont {Cooper}},
  \ and\ \bibinfo {author} {\bibfnamefont {J.~R.}\ \bibnamefont {Schrieffer}},\
  }\bibfield  {title} {\enquote {\bibinfo {title} {Theory of
  superconductivity},}\ }\href {\doibase 10.1103/PhysRev.108.1175} {\bibfield
  {journal} {\bibinfo  {journal} {Phys. Rev.}\ }\textbf {\bibinfo {volume}
  {108}},\ \bibinfo {pages} {1175} (\bibinfo {year}
  {1957}{\natexlab{b}})}\BibitemShut {NoStop}%
\bibitem [{\citenamefont {Ptok}\ and\ \citenamefont
  {Crivelli}(2017)}]{ptok.crivelli.17}%
  \BibitemOpen
  \bibfield  {author} {\bibinfo {author} {\bibfnamefont {A.}~\bibnamefont
  {Ptok}}\ and\ \bibinfo {author} {\bibfnamefont {D.}~\bibnamefont
  {Crivelli}},\ }\bibfield  {title} {\enquote {\bibinfo {title} {Influence of
  finite size effects on the {Fulde-Ferrell-Larkin-Ovchinnikov} state},}\
  }\href {\doibase 10.4208/cicp.OA-2016-0041} {\bibfield  {journal} {\bibinfo
  {journal} {Commun. Comput. Phys.}\ }\textbf {\bibinfo {volume} {21}},\
  \bibinfo {pages} {748} (\bibinfo {year} {2017})}\BibitemShut {NoStop}%
\bibitem [{\citenamefont {Januszewski}\ \emph {et~al.}(2015)\citenamefont
  {Januszewski}, \citenamefont {Ptok}, \citenamefont {Crivelli},\ and\
  \citenamefont {Gardas}}]{januszewski.ptok.15}%
  \BibitemOpen
  \bibfield  {author} {\bibinfo {author} {\bibfnamefont {M.}~\bibnamefont
  {Januszewski}}, \bibinfo {author} {\bibfnamefont {A.}~\bibnamefont {Ptok}},
  \bibinfo {author} {\bibfnamefont {D.}~\bibnamefont {Crivelli}}, \ and\
  \bibinfo {author} {\bibfnamefont {B.}~\bibnamefont {Gardas}},\ }\bibfield
  {title} {\enquote {\bibinfo {title} {{GPU}-based acceleration of free energy
  calculations in solid state physics},}\ }\href {\doibase
  10.1016/j.cpc.2015.02.012} {\bibfield  {journal} {\bibinfo  {journal}
  {Comput. Phys. Commun.}\ }\textbf {\bibinfo {volume} {192}},\ \bibinfo
  {pages} {220} (\bibinfo {year} {2015})}\BibitemShut {NoStop}%
\bibitem [{\citenamefont {Ma\'{s}ka}(1993)}]{maska.93}%
  \BibitemOpen
  \bibfield  {author} {\bibinfo {author} {\bibfnamefont {M.}~\bibnamefont
  {Ma\'{s}ka}},\ }\bibfield  {title} {\enquote {\bibinfo {title} {Self-energy
  approach to the {t-J} model},}\ }\href {\doibase 10.1103/PhysRevB.48.1160}
  {\bibfield  {journal} {\bibinfo  {journal} {Phys. Rev. B}\ }\textbf {\bibinfo
  {volume} {48}},\ \bibinfo {pages} {1160} (\bibinfo {year}
  {1993})}\BibitemShut {NoStop}%
\bibitem [{\citenamefont {Chen}\ \emph {et~al.}(2012)\citenamefont {Chen},
  \citenamefont {Gong},\ and\ \citenamefont {Zhang}}]{chen.gong.12}%
  \BibitemOpen
  \bibfield  {author} {\bibinfo {author} {\bibfnamefont {G.}~\bibnamefont
  {Chen}}, \bibinfo {author} {\bibfnamefont {M.}~\bibnamefont {Gong}}, \ and\
  \bibinfo {author} {\bibfnamefont {Ch.}\ \bibnamefont {Zhang}},\ }\bibfield
  {title} {\enquote {\bibinfo {title} {{BCS-BEC} crossover in
  spin-orbit-coupled two-dimensional {Fermi} gases},}\ }\href {\doibase
  10.1103/PhysRevA.85.013601} {\bibfield  {journal} {\bibinfo  {journal} {Phys.
  Rev. A}\ }\textbf {\bibinfo {volume} {85}},\ \bibinfo {pages} {013601}
  (\bibinfo {year} {2012})}\BibitemShut {NoStop}%
\bibitem [{\citenamefont {Shi}\ \emph {et~al.}(2016)\citenamefont {Shi},
  \citenamefont {Rosenberg}, \citenamefont {Chiesa},\ and\ \citenamefont
  {Zhang}}]{shi.rosenberg.16}%
  \BibitemOpen
  \bibfield  {author} {\bibinfo {author} {\bibfnamefont {H.}~\bibnamefont
  {Shi}}, \bibinfo {author} {\bibfnamefont {P.}~\bibnamefont {Rosenberg}},
  \bibinfo {author} {\bibfnamefont {S.}~\bibnamefont {Chiesa}}, \ and\ \bibinfo
  {author} {\bibfnamefont {S.}~\bibnamefont {Zhang}},\ }\bibfield  {title}
  {\enquote {\bibinfo {title} {Rashba spin-orbit coupling, strong interactions,
  and the {BCS-BEC} crossover in the ground state of the two-dimensional
  {Fermi} gas},}\ }\href {\doibase 10.1103/PhysRevLett.117.040401} {\bibfield
  {journal} {\bibinfo  {journal} {Phys. Rev. Lett.}\ }\textbf {\bibinfo
  {volume} {117}},\ \bibinfo {pages} {040401} (\bibinfo {year}
  {2016})}\BibitemShut {NoStop}%
\bibitem [{\citenamefont {Lee}\ and\ \citenamefont {Kim}(2017)}]{lee.kim.17}%
  \BibitemOpen
  \bibfield  {author} {\bibinfo {author} {\bibfnamefont {J.}~\bibnamefont
  {Lee}}\ and\ \bibinfo {author} {\bibfnamefont {D.-H.}\ \bibnamefont {Kim}},\
  }\bibfield  {title} {\enquote {\bibinfo {title} {Induced interactions in the
  {BCS-BEC} crossover of two-dimensional {Fermi} gases with {Rashba} spin-orbit
  coupling},}\ }\href {\doibase 10.1103/PhysRevA.95.033609} {\bibfield
  {journal} {\bibinfo  {journal} {Phys. Rev. A}\ }\textbf {\bibinfo {volume}
  {95}},\ \bibinfo {pages} {033609} (\bibinfo {year} {2017})}\BibitemShut
  {NoStop}%
\bibitem [{\citenamefont {Werthamer}\ \emph {et~al.}(1966)\citenamefont
  {Werthamer}, \citenamefont {Helfand},\ and\ \citenamefont {Hohenberg}}]{whh}%
  \BibitemOpen
  \bibfield  {author} {\bibinfo {author} {\bibfnamefont {N.~R.}\ \bibnamefont
  {Werthamer}}, \bibinfo {author} {\bibfnamefont {E.}~\bibnamefont {Helfand}},
  \ and\ \bibinfo {author} {\bibfnamefont {P.~C.}\ \bibnamefont {Hohenberg}},\
  }\bibfield  {title} {\enquote {\bibinfo {title} {Temperature and purity
  dependence of the superconducting critical field, {H$_{c2}$}. {III.} electron
  spin and spin-orbit effects},}\ }\href {\doibase 10.1103/PhysRev.147.295}
  {\bibfield  {journal} {\bibinfo  {journal} {Phys. Rev.}\ }\textbf {\bibinfo
  {volume} {147}},\ \bibinfo {pages} {295} (\bibinfo {year}
  {1966})}\BibitemShut {NoStop}%
\bibitem [{\citenamefont {Lei}\ \emph {et~al.}(2010)\citenamefont {Lei},
  \citenamefont {Hu}, \citenamefont {Choi}, \citenamefont {Warren},\ and\
  \citenamefont {Petrovic}}]{lei.hu.10}%
  \BibitemOpen
  \bibfield  {author} {\bibinfo {author} {\bibfnamefont {H.}~\bibnamefont
  {Lei}}, \bibinfo {author} {\bibfnamefont {R.}~\bibnamefont {Hu}}, \bibinfo
  {author} {\bibfnamefont {E.~S.}\ \bibnamefont {Choi}}, \bibinfo {author}
  {\bibfnamefont {J.~B.}\ \bibnamefont {Warren}}, \ and\ \bibinfo {author}
  {\bibfnamefont {C.}~\bibnamefont {Petrovic}},\ }\bibfield  {title} {\enquote
  {\bibinfo {title} {Pauli-limited upper critical field of
  {Fe$_{1+y}$Te$_{1-x}$Se$_{x}$}},}\ }\href {\doibase
  10.1103/PhysRevB.81.094518} {\bibfield  {journal} {\bibinfo  {journal} {Phys.
  Rev. B}\ }\textbf {\bibinfo {volume} {81}},\ \bibinfo {pages} {094518}
  (\bibinfo {year} {2010})}\BibitemShut {NoStop}%
\bibitem [{\citenamefont {Wolff-Fabris}\ \emph {et~al.}(2014)\citenamefont
  {Wolff-Fabris}, \citenamefont {Lei}, \citenamefont {Wosnitza},\ and\
  \citenamefont {Petrovic}}]{wolfffabris.lei.14}%
  \BibitemOpen
  \bibfield  {author} {\bibinfo {author} {\bibfnamefont {F.}~\bibnamefont
  {Wolff-Fabris}}, \bibinfo {author} {\bibfnamefont {H.}~\bibnamefont {Lei}},
  \bibinfo {author} {\bibfnamefont {J.}~\bibnamefont {Wosnitza}}, \ and\
  \bibinfo {author} {\bibfnamefont {C.}~\bibnamefont {Petrovic}},\ }\bibfield
  {title} {\enquote {\bibinfo {title} {Evolution of the pauli spin-paramagnetic
  effect on the upper critical fields of single-crystalline
  {K$_{x}$Fe$_{2-y}$Se$_{2-z}$S$_{z}$}},}\ }\href {\doibase
  10.1103/PhysRevB.90.024505} {\bibfield  {journal} {\bibinfo  {journal} {Phys.
  Rev. B}\ }\textbf {\bibinfo {volume} {90}},\ \bibinfo {pages} {024505}
  (\bibinfo {year} {2014})}\BibitemShut {NoStop}%
\bibitem [{\citenamefont {Weng}\ and\ \citenamefont {Hu}(2016)}]{weng.hu.16}%
  \BibitemOpen
  \bibfield  {author} {\bibinfo {author} {\bibfnamefont {K.-Ch.}\ \bibnamefont
  {Weng}}\ and\ \bibinfo {author} {\bibfnamefont {C.~D.}\ \bibnamefont {Hu}},\
  }\bibfield  {title} {\enquote {\bibinfo {title} {The p-wave superconductivity
  in the presence of {Rashba} interaction in {2DEG}},}\ }\href {\doibase
  10.1038/srep29919} {\bibfield  {journal} {\bibinfo  {journal} {Sci. Rep.}\
  }\textbf {\bibinfo {volume} {6}},\ \bibinfo {pages} {29919} (\bibinfo {year}
  {2016})}\BibitemShut {NoStop}%
\bibitem [{\citenamefont {Bloch}\ \emph {et~al.}(2008)\citenamefont {Bloch},
  \citenamefont {Dalibard},\ and\ \citenamefont {Zwerger}}]{bloch.dalibard.08}%
  \BibitemOpen
  \bibfield  {author} {\bibinfo {author} {\bibfnamefont {I.}~\bibnamefont
  {Bloch}}, \bibinfo {author} {\bibfnamefont {J.}~\bibnamefont {Dalibard}}, \
  and\ \bibinfo {author} {\bibfnamefont {W.}~\bibnamefont {Zwerger}},\
  }\bibfield  {title} {\enquote {\bibinfo {title} {Many-body physics with
  ultracold gases},}\ }\href {\doibase 10.1103/RevModPhys.80.885} {\bibfield
  {journal} {\bibinfo  {journal} {Rev. Mod. Phys.}\ }\textbf {\bibinfo {volume}
  {80}},\ \bibinfo {pages} {885} (\bibinfo {year} {2008})}\BibitemShut
  {NoStop}%
\bibitem [{\citenamefont {Jiang}\ \emph {et~al.}(2011)\citenamefont {Jiang},
  \citenamefont {Liu}, \citenamefont {Hu},\ and\ \citenamefont
  {Pu}}]{jiang.liu.11}%
  \BibitemOpen
  \bibfield  {author} {\bibinfo {author} {\bibfnamefont {L.}~\bibnamefont
  {Jiang}}, \bibinfo {author} {\bibfnamefont {X.-J.}\ \bibnamefont {Liu}},
  \bibinfo {author} {\bibfnamefont {H.}~\bibnamefont {Hu}}, \ and\ \bibinfo
  {author} {\bibfnamefont {H.}~\bibnamefont {Pu}},\ }\bibfield  {title}
  {\enquote {\bibinfo {title} {Rashba spin-orbit-coupled atomic {Fermi}
  gases},}\ }\href {\doibase 10.1103/PhysRevA.84.063618} {\bibfield  {journal}
  {\bibinfo  {journal} {Phys. Rev. A}\ }\textbf {\bibinfo {volume} {84}},\
  \bibinfo {pages} {063618} (\bibinfo {year} {2011})}\BibitemShut {NoStop}%
\bibitem [{\citenamefont {Galitski}\ and\ \citenamefont
  {Spielman}(2013)}]{galitski.spielman.13}%
  \BibitemOpen
  \bibfield  {author} {\bibinfo {author} {\bibfnamefont {V.}~\bibnamefont
  {Galitski}}\ and\ \bibinfo {author} {\bibfnamefont {I.~B.}\ \bibnamefont
  {Spielman}},\ }\bibfield  {title} {\enquote {\bibinfo {title} {Spin-orbit
  coupling in quantum gases},}\ }\href {\doibase 10.1038/nature11841}
  {\bibfield  {journal} {\bibinfo  {journal} {Nature}\ }\textbf {\bibinfo
  {volume} {494}},\ \bibinfo {pages} {49} (\bibinfo {year} {2013})}\BibitemShut
  {NoStop}%
\bibitem [{\citenamefont {Fu}\ \emph {et~al.}(2014)\citenamefont {Fu},
  \citenamefont {Huang}, \citenamefont {Meng}, \citenamefont {Wang},
  \citenamefont {Zhang}, \citenamefont {Zhang}, \citenamefont {Zhai},
  \citenamefont {Zhang},\ and\ \citenamefont {Zhang}}]{fu.huang.14}%
  \BibitemOpen
  \bibfield  {author} {\bibinfo {author} {\bibfnamefont {Z.}~\bibnamefont
  {Fu}}, \bibinfo {author} {\bibfnamefont {L.}~\bibnamefont {Huang}}, \bibinfo
  {author} {\bibfnamefont {Z.}~\bibnamefont {Meng}}, \bibinfo {author}
  {\bibfnamefont {P.}~\bibnamefont {Wang}}, \bibinfo {author} {\bibfnamefont
  {L.}~\bibnamefont {Zhang}}, \bibinfo {author} {\bibfnamefont
  {S.}~\bibnamefont {Zhang}}, \bibinfo {author} {\bibfnamefont
  {H.}~\bibnamefont {Zhai}}, \bibinfo {author} {\bibfnamefont {P.}~\bibnamefont
  {Zhang}}, \ and\ \bibinfo {author} {\bibfnamefont {J.}~\bibnamefont
  {Zhang}},\ }\bibfield  {title} {\enquote {\bibinfo {title} {Production of
  {Feshbach} molecules induced by spin-orbit coupling in {Fermi} gases},}\
  }\href {\doibase 10.1038/nphys2824} {\bibfield  {journal} {\bibinfo
  {journal} {Nat. Phys.}\ }\textbf {\bibinfo {volume} {10}},\ \bibinfo {pages}
  {110} (\bibinfo {year} {2014})}\BibitemShut {NoStop}%
\bibitem [{\citenamefont {Grusdt}\ \emph {et~al.}(2017)\citenamefont {Grusdt},
  \citenamefont {Li}, \citenamefont {Bloch},\ and\ \citenamefont
  {Demler}}]{grusdt.li.17}%
  \BibitemOpen
  \bibfield  {author} {\bibinfo {author} {\bibfnamefont {F.}~\bibnamefont
  {Grusdt}}, \bibinfo {author} {\bibfnamefont {T.}~\bibnamefont {Li}}, \bibinfo
  {author} {\bibfnamefont {I.}~\bibnamefont {Bloch}}, \ and\ \bibinfo {author}
  {\bibfnamefont {E.}~\bibnamefont {Demler}},\ }\bibfield  {title} {\enquote
  {\bibinfo {title} {Tunable spin-orbit coupling for ultracold atoms in
  two-dimensional optical lattices},}\ }\href {\doibase
  10.1103/PhysRevA.95.063617} {\bibfield  {journal} {\bibinfo  {journal} {Phys.
  Rev. A}\ }\textbf {\bibinfo {volume} {95}},\ \bibinfo {pages} {063617}
  (\bibinfo {year} {2017})}\BibitemShut {NoStop}%
\bibitem [{\citenamefont {Linder}\ and\ \citenamefont
  {Sudb\o{}}(2009)}]{linder.sudbo.09}%
  \BibitemOpen
  \bibfield  {author} {\bibinfo {author} {\bibfnamefont {J.}~\bibnamefont
  {Linder}}\ and\ \bibinfo {author} {\bibfnamefont {A.}~\bibnamefont
  {Sudb\o{}}},\ }\bibfield  {title} {\enquote {\bibinfo {title} {Theory of
  {Andreev} reflection in junctions with iron-based high-{T}$_{c}$
  superconductors},}\ }\href {\doibase 10.1103/PhysRevB.79.020501} {\bibfield
  {journal} {\bibinfo  {journal} {Phys. Rev. B}\ }\textbf {\bibinfo {volume}
  {79}},\ \bibinfo {pages} {020501} (\bibinfo {year} {2009})}\BibitemShut
  {NoStop}%
\end{thebibliography}%

\end{document}